\documentclass[aps,prd,amsmath,superscriptaddress,nofootinbib,%
               preprintnumbers,%
               onecolumn,12pt,tightenlines]{revtex4}

\usepackage{graphicx}

\usepackage{ifthen}

\graphicspath{{figures/}}

\newcommand{\coloronlinestatement}{}

\newcommand{\ie}{\textit{i.e.}}
\newcommand{\eg}{\textit{e.g.}}
\newcommand{\etal}{\textit{et~al.}}

\newcommand{\refeqn}[2][eqn:]{Eqn.~(\ref{#1#2})}

\newcommand{\reftab}[2][tab:]{Table~\ref{#1#2}}

\newcommand{\reffig}[2][fig:]{Figure~\ref{#1#2}}
\newcommand{\Reffig}[2][fig:]{Figure~\ref{#1#2}}
\newcommand{\refsec}[2][sec:]{Section~\ref{#1#2}} 

\newcommand{\ifmulticol}[2]{%
  \ifthenelse{\lengthtest{1.9\columnwidth<\textwidth}}{#1}{#2}%
}

\newcommand{\insertfig}[2][\scfigwidth]{%
    \hspace*{\stretch{1}}
    \includegraphics[keepaspectratio,width=#1\columnwidth]{#2}
    \hspace*{\stretch{1}}
}

\newcommand{\insertwidefig}[2][0.90]{%
    \hspace*{\stretch{1}}
    \includegraphics[keepaspectratio,width=#1\textwidth]{#2}
    \hspace*{\stretch{1}}
}

\newcommand{\insertdoublefig}[3][0.45]{%
    \hspace*{\stretch{1}}
    \includegraphics[keepaspectratio,width=#1\textwidth]{#2}
    \includegraphics[keepaspectratio,width=#1\textwidth]{#3}
    \hspace*{\stretch{1}}
}

\newcommand{\rmd}{\ensuremath{\mathrm{d}}}

\newcommand{\gae}{%
  \ensuremath{\lower 2pt \hbox{%
    $\, \buildrel {\scriptstyle >}\over {\scriptstyle \sim}\,$}%
    }%
  }
\newcommand{\lae}{%
  \ensuremath{\lower 2pt \hbox{%
    $\, \buildrel {\scriptstyle <}\over {\scriptstyle \sim}\,$}%
    }%
  }

\newcommand{\tanb}{\ensuremath{\tan{\beta}}}
\newcommand{\signmu}{\ensuremath{{\rm sign}(\mu)}}

\newcommand{\rmu}{\ensuremath{\mathrm{u}}}
\newcommand{\rms}{\ensuremath{\mathrm{s}}}
\newcommand{\rmc}{\ensuremath{\mathrm{c}}}
\newcommand{\rmb}{\ensuremath{\mathrm{b}}}
\newcommand{\rmt}{\ensuremath{\mathrm{t}}}
\newcommand{\rmpp}{\ensuremath{\mathrm{p}}}
\newcommand{\rmnn}{\ensuremath{\mathrm{n}}}

\newcommand{\mup}{\ensuremath{m_{\rm u}}}
\newcommand{\mumd}{\ensuremath{m_{\rm u} / m_{\rm d}}}
\newcommand{\md}{\ensuremath{m_{\rm d}}}
\newcommand{\ms}{\ensuremath{m_{\rm s}}}
\newcommand{\msmd}{\ensuremath{m_{\rm s} / m_{\rm d}}}
\newcommand{\mc}{\ensuremath{m_{\rm c}}}
\newcommand{\mb}{\ensuremath{m_{\rm b}}}
\newcommand{\mt}{\ensuremath{m_{\rm t}}}
\newcommand{\SigmapiN}{\ensuremath{\Sigma_{\pi\!{\scriptscriptstyle N}}}}
\newcommand{\athree}{\ensuremath{a_{3}^{\rm (p)}}}
\newcommand{\aeight}{\ensuremath{a_{8}^{\rm (p)}}}
\newcommand{\Deltaps}{\ensuremath{\Delta_{\rms}^{(\rmpp)}}}

\newcommand{\rhoDM}{\ensuremath{\rho_{\chi}}}
\newcommand{\fTq}[1]{\ensuremath{f_{T_{#1}}}}
\newcommand{\fNTq}[1]{\ensuremath{f_{T_{#1}}^{(N)}}}

\newcommand{\Bq}[1]{\ensuremath{B_{#1}}}
\newcommand{\BNq}[1]{\ensuremath{B_{#1}^{(N)}}}
\newcommand{\Bpq}[1]{\ensuremath{B_{#1}^{(\rmpp)}}}
\newcommand{\Bnq}[1]{\ensuremath{B_{#1}^{(\rmnn)}}}

\newcommand{\DeltaNq}[1]{\ensuremath{\Delta_{#1}^{(N)}}}
\newcommand{\Deltapq}[1]{\ensuremath{\Delta_{#1}^{(\rmpp)}}}
\newcommand{\Deltanq}[1]{\ensuremath{\Delta_{#1}^{(\rmnn)}}}
\newcommand{\Eth}{\ensuremath{E_{\rm th}}}

\newcommand{\betap}{\ensuremath{\beta_{+}}}
\newcommand{\betam}{\ensuremath{\beta_{-}}}
\newcommand{\GG}{\ensuremath{\mathcal{G}}}

\begin{document}


\preprint{CERN-PH-TH/2009-245}
\preprint{UMN--TH--2829/09}
\preprint{FTPI--MINN--09/47}


\title{Neutrino Fluxes from CMSSM LSP Annihilations in the Sun}

\author{\bf John Ellis}
\email{John.Ellis@cern.ch}
\affiliation{
 TH Division, Physics Department,
 CERN,
 1211 Geneva 23, Switzerland}

\author{\bf Keith A. Olive}
\email{olive@physics.umn.edu}
\affiliation{
 William I. Fine Theoretical Physics Institute,
 School of Physics and Astronomy,
 University of Minnesota,
 Minneapolis, MN 55455, USA}

\author{\bf Christopher Savage}
\email{savage@fysik.su.se}
\affiliation{
 The Oskar Klein Centre for Cosmoparticle Physics,
 Department of Physics,
 Stockholm University,
 AlbaNova,
 SE-10691 Stockholm, Sweden}

\author{\bf Vassilis C. Spanos}
\email{vspanos@physics.upatras.gr}
\affiliation{
 Department of Physics,
 University of Patras,
 GR-26500 Patras, Greece}

\date{March 3, 2010}        


\begin{abstract}
\vskip .2in

\centerline{\bf Abstract}

\vskip .2in

{\small We evaluate the neutrino fluxes to be expected from
neutralino LSP annihilations inside the Sun, within the
minimal supersymmetric extension of the Standard Model with
supersymmetry-breaking scalar and gaugino masses constrained
to be universal at the GUT scale (the CMSSM). We find that there are 
large regions of typical CMSSM $(m_{1/2}, m_0)$ planes where the LSP
density inside the Sun is not in equilibrium, 
so that the annihilation rate may be far below the capture rate.
We show that neutrino fluxes are dependent on the
solar model at the 20\% level, and adopt the AGSS09 model of Serenelli
{\it et~al.}  for our detailed studies. We find that there are large
regions of the CMSSM $(m_{1/2}, m_0)$ planes where the capture rate is
not dominated by spin-dependent LSP-proton scattering, e.g., at large
$m_{1/2}$ along the CMSSM coannihilation strip. We calculate neutrino
fluxes above various threshold energies for points along the
coannihilation/rapid-annihilation and focus-point strips 
where the CMSSM yields the correct cosmological relic density for
$\tanb = 10$ and 55 for $\mu > 0$, exploring their sensitivities to
uncertainties in the spin-dependent and -independent scattering matrix
elements. We also present detailed neutrino
spectra for four benchmark models that illustrate generic possibilities
within the CMSSM.
Scanning the cosmologically-favored parts of the parameter space of the
CMSSM, we find that the IceCube/DeepCore detector can probe at best only
parts of this parameter space, notably the focus-point region
and possibly also at the low-mass tip of the coannihilation strip.
}

\end{abstract}

\maketitle

\vfill
\leftline{CERN-PH-TH/2009-245}
\leftline{December  2009}


\section{\label{sec:intro} Introduction}

The most convincing way  to verify directly the existence of
astrophysical cold dark matter particles would be through their
scattering on nuclei in low-background underground
experiments~\cite{Goodman:1984dc}.  Complementing this search for dark
matter scattering in the laboratory via spin-independent and/or
-dependent scattering, the next most direct way to search to confirm
the nature of any such dark matter particles would be to observe
products of their annihilations in an astrophysical context.
Although less direct, observations of such annihilation products would
provide valuable insight into the annihilation rates into important
channels, which would provide more information on the dynamics of the
dark matter particles, beyond their scattering cross sections.
Observations of astrophysical annihilations of dark matter particles
would be particularly interesting because cosmological annihilations 
earlier in the history of the Universe controlled the primordial relic
abundance.

Several environments for annihilations of astrophysical dark matter
particles are of interest, including
the galactic halo \cite{indirectdet:galactichalo},
the galactic centre \cite{indirectdet:galacticcenter},
the Sun \cite{indirectdet:solar},
the Earth \cite{indirectdet:earth},
dwarf galaxies \cite{indirect:dwarfs} 
and galaxy clusters \cite{indirect:clusters}:
our focus here is on the flux of neutralinos from dark matter
annihilations inside the Sun.  In this case, the dominant astrophysical
uncertainties are the total line density of cold dark matter through
which the Sun has passed throughout its history, and the mass fractions
of different elements in the Sun, particularly for heavy nuclei (metals).
In this paper we do not discuss the former, 
but we compare the predictions of models with
different element compositions and discuss the importance of this
model dependence for the observability of a signal.

There have been many studies of the prospective neutrino fluxes
from solar dark matter annihilations in generic versions of the Minimal
Supersymmetric extension of the Standard Model (MSSM), and also in
specific versions such as the CMSSM \cite{cmssmnu}, in which
the supersymmetry-breaking scalar and gaugino masses are each
constrained to be universal at some input GUT scale, and the lightest 
neutralino, $\chi$, is assumed to be the lightest supersymmetric
particle (LSP) \cite{cmssm,eoss,cmssmwmap}. 
In this paper we study $\chi$ annihilations in the CMSSM, focusing
in particular on the strips of parameter space \cite{eoss} where the
relic $\chi$ density lies within the range of cold dark matter density
indicated by WMAP \cite{Komatsu:2008hk} and other 
cosmological measurements. We also present more complete results for
a few specific benchmark CMSSM scenarios \cite{bench}, which are
representative of the range of possibilities within the CMSSM.

The first step in calculating the flux of neutrinos from LSP
annihilations inside the Sun is to calculate the LSP capture rate.
This is controlled by the cross sections for LSP scattering on
protons, which is mainly spin-dependent in the CMSSM, and on heavier
nuclei, which is mainly spin-independent.
Early work on capture often assumed that spin-dependent scattering would
dominate, and neglected the spin-independent contribution from heavier
nuclei, in particular.  Secondly, one must check whether it is
correct to assume that the LSP capture
and annihilation processes in the Sun have reached equilibrium.

As we show later in this paper, neither of these assumptions is valid in
general in the CMSSM: spin-independent scattering is also important,
particularly at large values of $m_{1/2}$, and the annihilation rate is 
significantly smaller than the capture rate in large areas
of the $(m_{1/2}, m_0)$ plane. Both these effects are particularly 
important at large $m_{1/2}$ along the coannihilation WMAP strip.
We also discuss the differences in the annihilation rates calculated using
different solar models.

The spin-dependent and -independent dark matter scattering rates both
have significant uncertainties that affect the capture rate and hence
also the annihilation rate \cite{Bottino:1999ei,Accomando:1999eg,
Ellis:2005mb,Ellis:2008hf,Niro:2009mw}.
The matrix element for spin-dependent scattering on a proton is related
to the decomposition of the proton spin, in which the greatest
uncertainty is the contribution of the strange 
quarks and antiquarks, $\Deltaps$. Inclusive deep-inelastic scattering
experiments favour $\Deltaps = - 0.09 \pm 0.03$ \cite{Alekseev:2007vi}, 
whereas analyses of
particle production in deep-inelastic scattering are quite compatible
with $\Deltaps = 0$~\cite{Airapetian:2008qf}. In this paper, we compare the
neutrino fluxes expected for $\Deltaps = 0, -0.06, -0.09$ and $- 0.12$. 

On the other hand, the matrix
elements for spin-independent scattering on nuclei are related to the
$\pi$-nucleon $\sigma$ term, $\SigmapiN$, whose value is also quite
uncertain. The central value in a theoretical analysis of low-energy
$\pi$-nucleon scattering data is $\SigmapiN = 64$~MeV
\cite{Pavan:2001wz}, whereas $\SigmapiN = 36$~MeV would correspond to
the absence of a strange scalar density in the nucleon:
$\langle N | {\bar s}s | N \rangle = 0$. We consider both these values,
as well as the intermediate value $\SigmapiN = 45$~MeV. Lattice
calculations are now reaching the stage where they may also provide 
useful information on $\SigmapiN$~\cite{Young:2009zb}, and a recent
analysis would suggest a lower value $\SigmapiN \lae 40$ \cite{Giedt:2009mr}.
For comparison, we also calculate the flux generated
if spin-independent scattering is neglected altogether.

The layout of this paper is as follows. In \refsec{cmssm} we discuss relevant
aspects of the CMSSM, introducing general aspects of the $(m_{1/2}, m_0)$
planes for $\tanb = 10$ and 55 that we use as the basis for our
subsequent calculations. In \refsec{astrophysical} we present various
astrophysical considerations, including a comparison of the LSP
capture and annihilation rates in the CMSSM, and a discussion of
sensitivity to the solar model. In \refsec{hadronic} we discuss the
sensitivities of the annihilation rates to uncertainties in the Standard
Model, principally the hadronic matrix elements $\Deltaps = 0$ and
$\SigmapiN$ that control the spin-dependent and -independent scattering
LSP-matter scattering cross sections, respectively, both of  which are
relevant, and the latter quite important. Then, in \refsec{fluxes} we
discuss the neutrino and neutrino-induced muon fluxes from LSP
annihilations, first over the $(m_{1/2}, m_0)$
planes for $\tanb = 10$ and 55, subsequently along the WMAP
strips in the coannihilation/rapid-annihilation and focus-point strips,
and finally for four specific benchmark scenarios. We present results
for three different neutrino thresholds: 1, 10 and 100~GeV.
We find that restricted portions of the WMAP strips yields signals that may be
detectable in the IceCube/DeepCore experiment \cite{Ahrens:2003ix,
Wiebusch:2009jf}.
In the coannihilation region, only models with small $m_{1/2}$
are detectable in IceCube/DeepCore.
In the case of the focus-point region, a more extended part of the strip
may be detectable in IceCube/DeepCore for $\tanb = 10$.
For $\tanb = 55$, even in the focus-point region only models with small
$m_{1/2}$ may be detectable in IceCube/DeepCore.
These conclusions are not very sensitive to the uncertainties in the
solar model and in $\Deltaps = 0$ that
controls spin-dependent scattering, but the
range of CMSSM parameter space that might be
observable depends on the hadronic matrix element
$\SigmapiN$ controlling spin-independent scattering.

\section{\label{sec:cmssm} CMSSM Parameter Space}

We set the scene by first discussing the CMSSM parameter space
that we explore. Points in the CMSSM are in principle characterized
by four free parameters, the common gaugino mass, $m_{1/2}$, the
common scalar mass, $m_0$, the common trilinear supersymmetry-breaking
parameter, $A_0$, and the ratio of Higgs vacuum expectation values,
$\tanb$.  There is also an ambiguity in the sign of the Higgs mixing
parameter, $\mu$: motivated by $g_\mu - 2$ and $b \to s \gamma$, we
restrict our attention to positive $\mu$~\footnote{
  We neglect the possibility of significant CP-violating
  phases in the soft supersymmetry-breaking parameters.
  }.
The three most relevant parameters for this analysis are $m_{1/2}, m_0$
and $\tanb$, and we present our results in the $(m_{1/2}, m_0)$
planes for the trilinear soft supersymmetry-breaking parameter
$A_0 = 0$ and two discrete choices $\tanb = 10, 55$. The
choice $\tanb = 10$ is close to the optimal value we find in a global
likelihood analysis incorporating all the theoretical,
phenomenological, experimental and cosmological constraints on the
CMSSM parameter space \cite{mc3}. The choice $\tanb = 55$ is close the
maximum value for which we find consistent solutions of the CMSSM vacuum
conditions, and represents the most distinct alternative from the
favoured case $\tanb = 10$.

\begin{figure*}
  \insertdoublefig{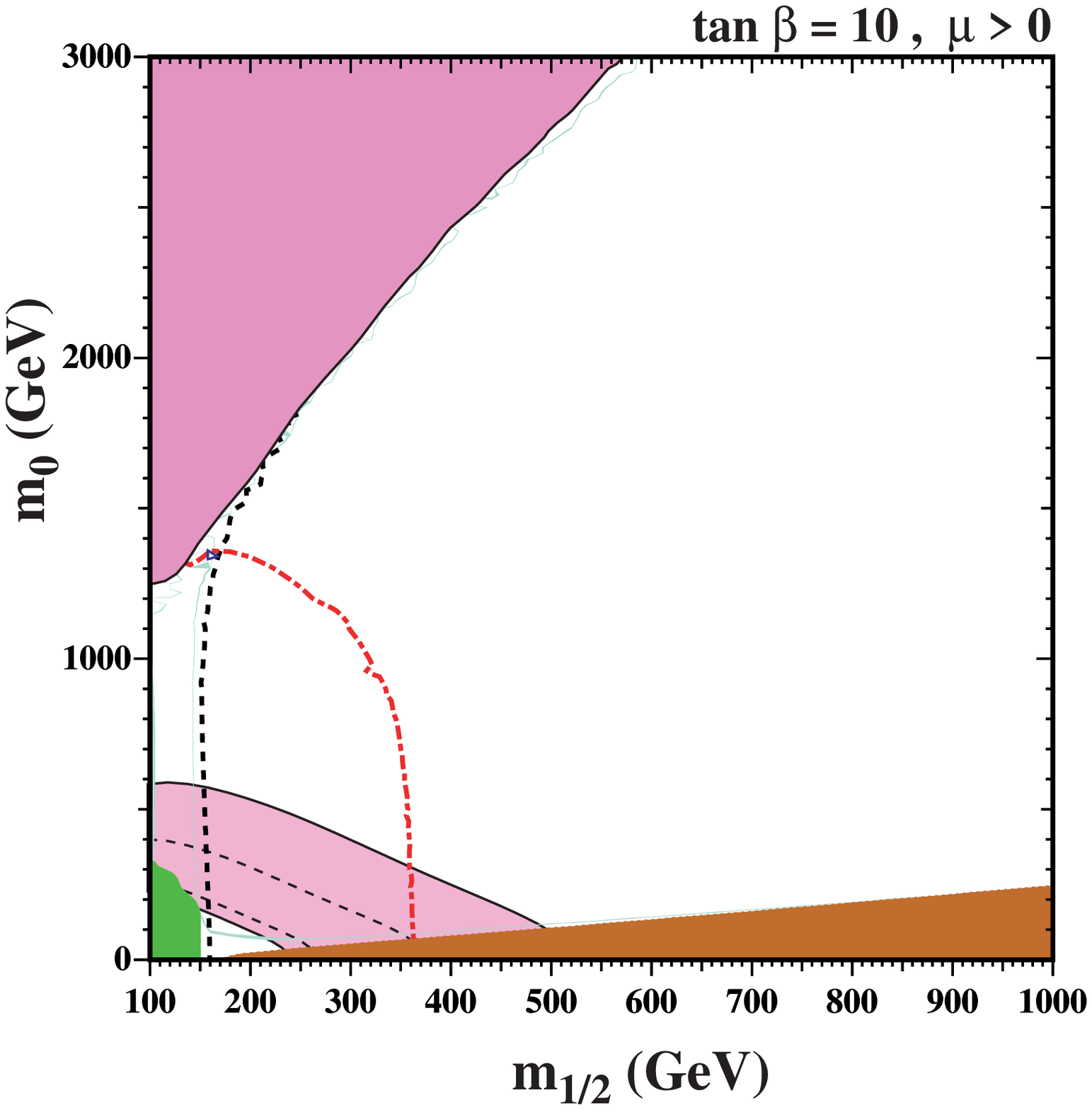}{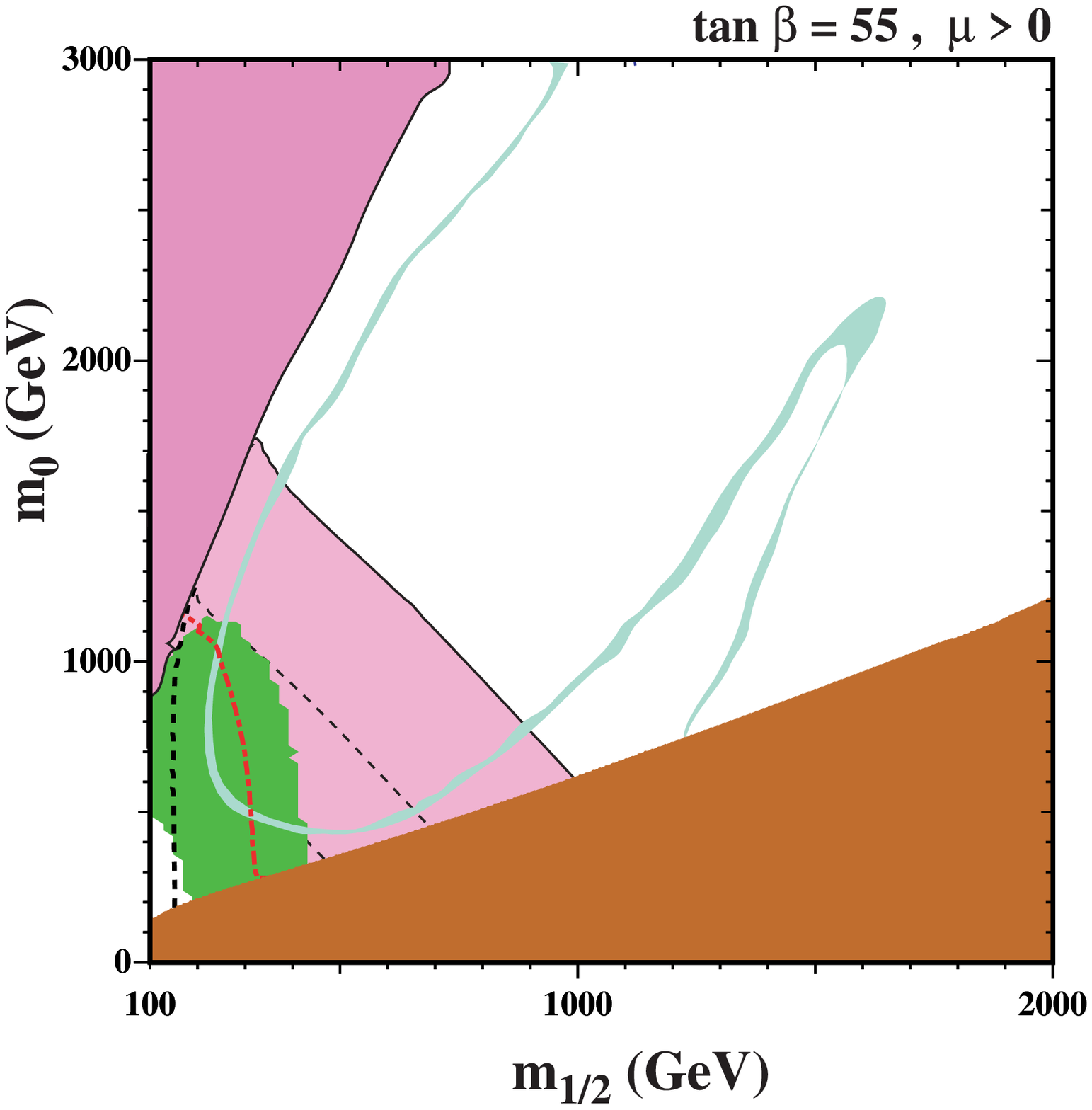}
  \caption{\it
    \coloronlinestatement
    The $(m_{1/2},m_0)$ planes for the CMSSM for $A_0 = 0$ and (left) 
    $\tanb = 10$, (right) $\tanb = 55$, showing regions excluded
    because there is no consistent electroweak vacuum (dark pink
    shading), or because there is charged dark matter (brown shading),
    or because of a conflict with $b \to s \gamma$ measurements (green
    shading).  Only regions to the right of the black dashed (red
    dash-dotted) line are consistent with the absence at LEP of
    charginos (Higgs boson).  The turquoise strips are favoured by the
    determination of the cold dark matter density by WMAP and other
    experiments \cite{Komatsu:2008hk}, and the pink strips are favoured
    by the BNL measurement of $g_\mu - 2$.
  }
  \label{fig:planes}
\end{figure*}

In \reffig{planes} and subsequent figures we display these
$(m_{1/2}, m_0)$ planes with the following constraints implemented:
regions lacking a consistent electroweak vacuum are found in the upper
left corners of the figures (dark pink shading),
regions with charged dark matter  are found in the lower right corners
(brown shading),
the LEP bound on charginos~\cite{LEPsusy} excludes the area to the left
of black dashed line and
the LEP bound of 114.4 GeV on the Higgs mass~\cite{LEPHiggs}
is shown by (red) dash-dotted line (lighter Higgs masses
occur to the left of this line). 
Here the code {\tt FeynHiggs}~\cite{FeynHiggs} is used for the
calculation of $m_h$.
In addition we impose agreement with $b \to s \gamma$
measurements~\cite{bsgex} which excludes the green shaded region at
relatively low $m_{1/2}$.
We also display the strips favoured by the determination
of the cold dark matter density by WMAP and other experiments
\cite{Komatsu:2008hk} (turquoise shading). Also shown (shaded pink) is
the band favoured by the BNL measurement of $g_\mu - 2$~\cite{Bennett:2006fi},
using the latest estimate of the Standard Model contribution based on a
compilation of $e^+ e^-$ data including the most recent
{\it B{\small A}B{\small AR}} result~\cite{g-2babar}, which leaves a
discrepancy
$\Delta (g_\mu - 2)/2 = (24.6 \pm 8.0) \times 10^{-10}$~\cite{newg-2}
that could be explained by supersymmetry~\footnote{
  The dashed lines include the 1-$\sigma$ range
  of $\Delta (g_\mu - 2)/2$, the solid lines the 2-$\sigma$ range.}.

In the case $\tanb = 10$ (left panel of \reffig{planes}), there are two
WMAP strips, one
close to the boundary with charged dark matter where the neutralino
coannihilates with the lighter stau and other sleptons, and the other
close to the boundary of the region with an inconsistent electroweak
vacuum, the focus-point region. In the case $\tanb = 55$ (right panel
of \reffig{planes}), the focus-point strip moves away from the
electroweak vacuum boundary at large $m_0$ and
small $m_{1/2}$, and connects with the coannihilation strip.
The latter moves away from the charged dark matter region at small
$m_0$ and large $m_{1/2}$, and bifurcates into strips on either
side of a funnel where neutralinos annihilate rapidly via
direct-channel $H, A$ poles.

In regions between the WMAP strips and the boundaries set by the
absence of charged dark matter and the consistency of the electroweak
vacuum, the calculated relic density falls below the WMAP range
\cite{Komatsu:2008hk}. In
these regions, for the purposes of our subsequent calculations we
rescale the halo density by the ratio of the calculated cosmological
density to the 2-$\sigma$ lower limit of the WMAP range. The same scaling
is applied between the two sides of the funnel when $\tanb = 55$.
Elsewhere, between the
WMAP strips, the calculated density exceeds the WMAP range
if one assumes conventional cosmology, but we
nevertheless assume naively that the halo density is saturated by relic
neutralinos. This would be possible, in principle, if some mechanism
increased the entropy density between the epochs of dark matter
freeze-out, when the temperature $T \sim m_\chi/20 \gae 5$~GeV, and of
cosmological nucleosythesis, when $T \lae 1$~MeV. We do not specifically
advocate such a scenario, but displaying complete $(m_{1/2}, m_0)$
planes enables us to put in context the results we present later for
models along the
WMAP strips. We also present more detailed results for representative
benchmark points located on these strips.

\section{\label{sec:astrophysical} Astrophysical Considerations}

We review here the basic capture and annihilation processes and
astrophysical considerations that affect these processes.
The capture and annihilation of LSPs in the Sun depends both on the
distribution of LSPs in the halo and on the abundances and distribution
of elements within the Sun.

For the dark matter halo, we take a non-rotating
isothermal sphere \cite{Freese:1987wu} with an rms speed of 270~km/s,
a disk rotation speed of 220~km/s, and a local dark matter density of
0.3~GeV/cm$^3$,
and we do not address other halo models in this paper. 
Our results would scale linearly with the local dark matter density as
equilibrium between capture and annihilation is approached.
For our fiducial solar model, we use the AGSS09 model of Serenelli
\etal\ \cite{Serenelli:2009yc}, which is based
on the proto-solar isotopic abundances found in Ref.~\cite{Asplund:2009fu}.

Gravitational capture of LSPs occurs when LSPs from the galaxy pass
through the Sun, scatter off a nucleus, and lose enough energy that
they can no longer escape from the Sun's gravitational potential.  The
capture rate per unit stellar mass via scattering off an isotope
indexed by $i$ is given by \cite{Gould:1987ir,Gould:1992xx}:
\begin{equation} \label{eqn:dCidM}
  \frac{\rmd C_i}{\rmd M}(v) =
    \frac{\rhoDM \sigma_i}{m_{\chi} m_i} \, \epsilon_i \,
    \int \rmd^3u \, \frac{f(u)}{u}
    \, (u^2 + v^2) \, \GG_i(u,v) \, ,
\end{equation}
where $\rhoDM$ is the LSP density,
$m_{\chi}$ is the LSP mass, $m_i$ is the nuclear mass,
$\sigma_i$ is the LSP-nuclear scattering cross-section at zero
momentum transfer, $\epsilon_i$ is the mass fraction of the isotope,
$u$ is the LSP's velocity far away from the Sun, $v(r)$ is the 
velocity required for escape from the Sun, and
\begin{equation} \label{eqn:fu}
  f(u) \equiv \frac{1}{4 \pi} \int \rmd\Omega \, f(\mathbf{u})
\end{equation}
is the angular-averaged LSP velocity distribution in the galactic
neighborhood (in the Sun's frame, but outside the solar potential).
Here,
\begin{equation} \label{eqn:Gi}
  \GG_i(u,v) \equiv
    \frac{1}{E_{\max}}
    \int_{m_{\chi} u^2/2}^{E_{\max}}
    \rmd(\Delta E) \, |F_i(\Delta E)|^2
    \, \theta(\betam v^2 - u^2) \, ,
\end{equation}
where
\begin{equation} \label{eqn:beta}
  \beta_{\pm} \equiv \frac{4 m_{\chi} m_i}{(m_{\chi} \pm m_i)^2} \, ,
\end{equation}
$|F_i(\Delta E)|^2$ is a form factor (discussed below),
$\Delta E$ is the energy imparted to a stationary nucleus in a
collision with an LSP,
and $E_{\max} = \frac{1}{2} \betap m_{\chi} (u^2+v^2)$ is the maximum
energy that can be imparted
by an LSP with velocity $\sqrt{u^2+v^2}$, where $\sqrt{u^2+v^2}$ is
the velocity of the LSP in the Sun.  The quantity $\GG_i(u,v)$
factors in two effects: (1) only a fraction of the scatters result in
LSPs losing enough energy to become gravitationally bound, and
(2) a form factor for finite momentum exchange can suppress
higher-energy collisions.
In the absence of a form factor ($|F_i|^2 = 1$), $\GG_i(u,v)$ is just
the fraction of scatters of an LSP at velocity $\sqrt{u^2+v^2}$
on a stationary nucleus that yield a gravitationally-bound LSP.
The theta function $\theta(\betam v^2 - u^2)$ accounts for the fact
that, for sufficiently large LSP velocities, scattering into bound
orbits is not kinematically possible.
The total capture rate is obtained by integrating \refeqn{dCidM}
over the solar profile and summing over the isotopes present in the
Sun.

The derivation of \refeqn{dCidM} is lengthy and the equations resulting
from carrying out the integration are long and uniformative, so we do
not include them here.  Both the source of \refeqn{dCidM} and
integrations are given by Gould in Refs.~\cite{Gould:1987ir}
\& \cite{Gould:1992xx}.  The capture rate is arrived at by different
methods in those two references; our forms above closely match those
of the latter, albeit with some small differences~\footnote{
  The quantity $\GG(u,v)$ is equivalent to
  $\frac{1}{E_{\max}} \, G(u,v)$ in Ref.~\cite{Gould:1992xx},
  whilst the LSP-nuclear scattering cross-section $\sigma_i$ is
  equivalent to the quantity $\betap m_{\chi} m_i \sigma_0 \, Q^2$.
  We also differ from Gould in that $f(u)$ is both the spatial and
  velocity phase space density in his derivations
  ($\int \rmd^3u f(u) = \rhoDM/m_{\chi}$), as opposed to just the
  velocity phase space density here ($\int \rmd^3u f(u) = 1$).
  }.
We refer the reader to those references for technical details and a more
thorough discussion of the capture process.  Here, we
point out some of the features of \refeqn{dCidM} and \refeqn{Gi}:
\begin{itemize}
  \item The fraction of scatters that yield capture is largest
        for nuclei with masses nearest that of the LSP.  Capture
        is suppressed for scattering off of light nuclei
        ($m_i \ll m_{\chi}$) as the LSP loses very little energy in
        these collisions and is more likely to maintain enough speed to
        escape the solar system and continue back out into the halo.
  \item The fraction of scatters that yield capture is higher for
        scatters that occur deeper in the solar potential (larger $v$),
        \ie\ when closer to the Sun's center.
  \item Slow LSPs (small $u$) require less of an energy loss to become
        gravitationally bound and are more easily captured.
\end{itemize}

Scattering in the Sun is complicated by the fact that some LSP-nuclear
interactions, particularly those involving heavy nuclei, are energetic
enough that the finite size of the nucleus becomes important.
In the zero-momentum-transfer limit, the LSP is expected to scatter
coherently off the entire nucleus, which can effectively be treated
as a point particle; an isotropic scattering cross section $\sigma_i$
is assumed in this case.  When the inverse of the momentum $q$
exchanged in the scatter becomes comparable to or smaller than the
size of the nucleus, the point-particle approximation is no longer valid
and we must include a form factor: $\sigma_i \to \sigma_i |F(q^2)|^2$.
We use simple exponential form factors,
\begin{equation} \label{eqn:FF}
  |F(\Delta E)|^2 = e^{-\Delta E/E_0} \, ,
  \qquad\qquad
  E_0 = \begin{cases}
          \frac{3 \hbar^2}{2 m_i r_i^2} & \quad\textrm{(spin-dependent)} \\
          \frac{5 \hbar^2}{2 m_i r_i^2} & \quad\textrm{(spin-independent)}
        \end{cases}
\end{equation}
where $m_i$ is the nuclear mass, $r_i$ is the effective nuclear radius,
and $\Delta E = \frac{q^2}{2M}$ is the energy lost by the LSP in the
collision.  Capture rates with a form factor of this form can be
found in the Appendix of Ref.~\cite{Gould:1987ir}.  We differ here
from Gould and other determinations of capture rates in that we use
a different scale $E_0$ in the exponential for spin-dependent
and spin-independent scattering \footnote{
  Note that Gould \cite{Gould:1987ir,Gould:1992xx} 
  defines the exponential scale in terms of the rms
  radius of the nucleus instead of the effective radius used here.
  }.
For the latter case, this is a
reasonable approximation to the more accurate Helm form factor
for $\Delta E \lae 2 E_0$ \cite{SmithLewin,Duda:2006uk}.
While the exponential form here becomes a poor approximation to the true
form factor when $\Delta E \gg E_0$, the number of scattering events
at such large energy losses are suppressed to the point that they make
little contribution to the overall capture rate.  The capture rate is
thus sensitive mainly to the energy at which the form factor becomes
important (\ie\ $E_0$), not the shape of the form factor beyond this
energy.  The exponential form factor we use is therefore a reasonable
approximation and use of \eg\ the Helm form factor is not necessary
for determining capture rates.
This differs from direct detection searches, where the signal may be
mainly high-energy scatters, for which a form factor that is accurate at
large $\Delta E$ is necessary.

As gravitationally-bound LSPs undergo additional scatters on subsequent
passes through the Sun, they fall inwards and approach
thermal equilibrium at the center of the Sun.  As the population of
LSPs increases, the rate at which they annihilate with each other
also increases; given sufficient time, the annihilation rate will come
to equilibrium with the rate at which the LSPs are captured.
The number $N$ of thermalized LSPs follows
\begin{equation} \label{eqn:Ndiffeq}
  \frac{\rmd N}{\rmd t} = C - C_E N - C_A N^2 \, ,
\end{equation}
where $C$ is the capture rate, $C_E$ parametrizes the
evaporation rate of the thermalized LSPs, and $C_A$ parametrizes
the annihilation rate of the LSPs.  Evaporation is only significant
for LSPs lighter than $\sim$3~GeV \cite{Gould:1987ju} and is
neglected here, as we do not examine any models with such
light LSPs.
To a good approximation,
\begin{equation} \label{eqn:CA}
  C_A = <\sigma_A v> \left( \frac{3 k T_c}{2 G m_{\chi} \rho_c} \right)^{-3/2}
  \, ,
\end{equation}
where 
$<\sigma_A v>$ is the total LSP s-wave annihilation cross section,
$m_{\chi}$ is the LSP mass, and $T_c$ and
$\rho_c$ are the temperature and density, respectively, at the center
of the Sun.  In the AGSS09 model, $T_c$ is $1.55 \times 10^7$~K and
$\rho_c$ is 151~g/cm$^3$; these two quantities vary by negligible
amounts for other solar models.
Taking $C_E = 0$ and solving \refeqn{Ndiffeq} yields an annihilation
rate of
\begin{equation} \label{eqn:ar}
  2 \Gamma = C \tanh^2\left(\frac{t_{\odot}}{\tau} \right) \, ,
\end{equation}
where $t_{\odot}$ is the age of the Sun and $\tau = 1/\sqrt{C C_A}$.
The factor of 2 in $2 \Gamma$ is due to the loss of two LSPs in each
annihilation event.

It is often assumed that, after the first scatter in which an LSP
becomes gravitationally bound, it will quickly undergo additional
scatters and come to thermal equilibrium in the center of the Sun.
This assumption is not always valid: in some cases, further scatters
may occur over time periods longer than the age of the Sun, in which
case the LSPs have not yet reached thermal equilibrium, or the initial
orbits may extend far enough out into the solar system that the LSPs
are perturbed or gravitationally scattered by the planets, in which
case they will never come to thermal
equilibrium~\cite{Savage:2007aa,Peter:2009mk}.
In both cases, the annihilation
rate will be suppressed.  However, for the CMSSM regions examined here,
these effects are negligible and we ignore them.

\subsection{Comparison of Capture and Annihilation Rates}

\begin{figure*}
  \insertdoublefig{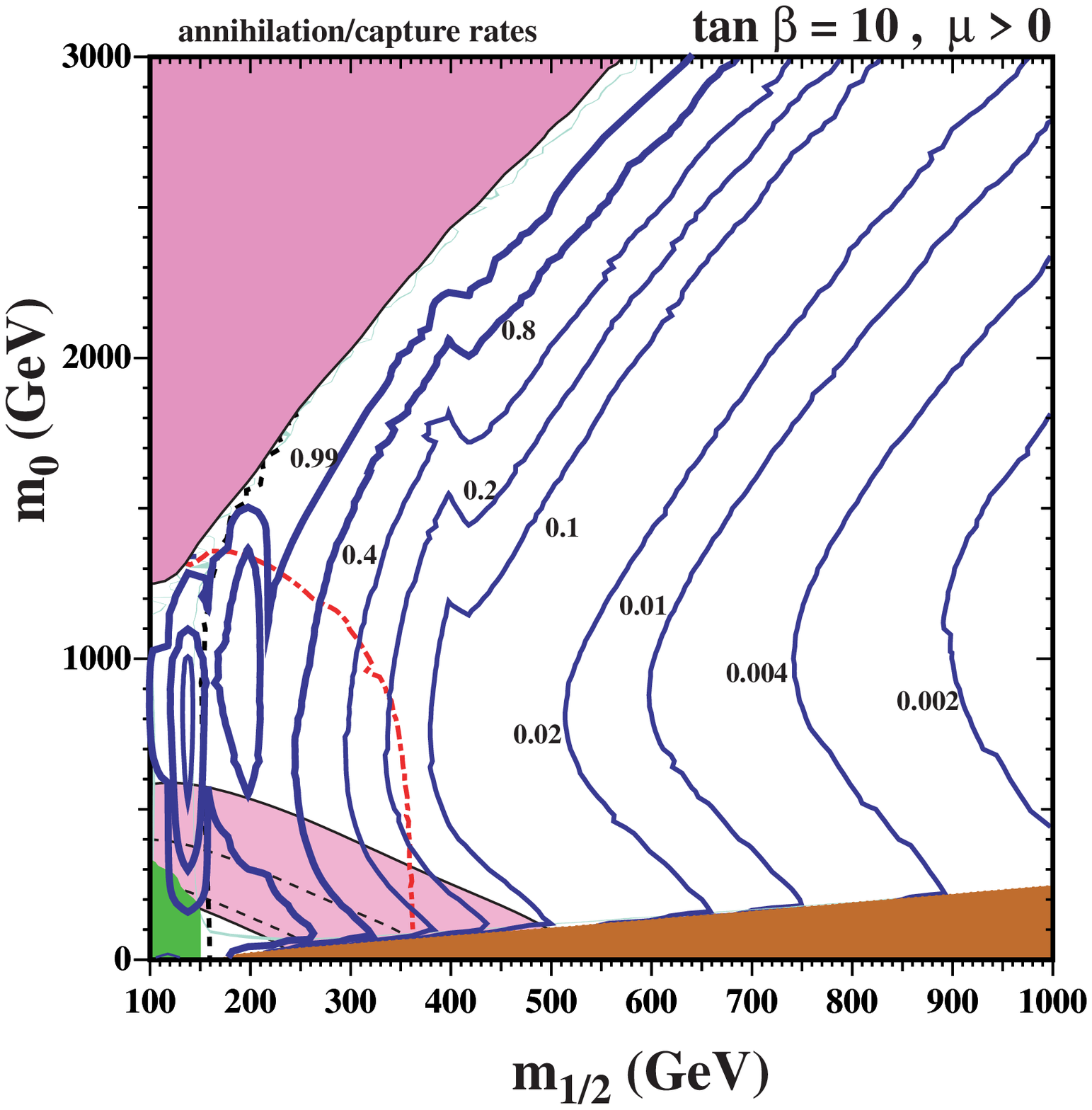}{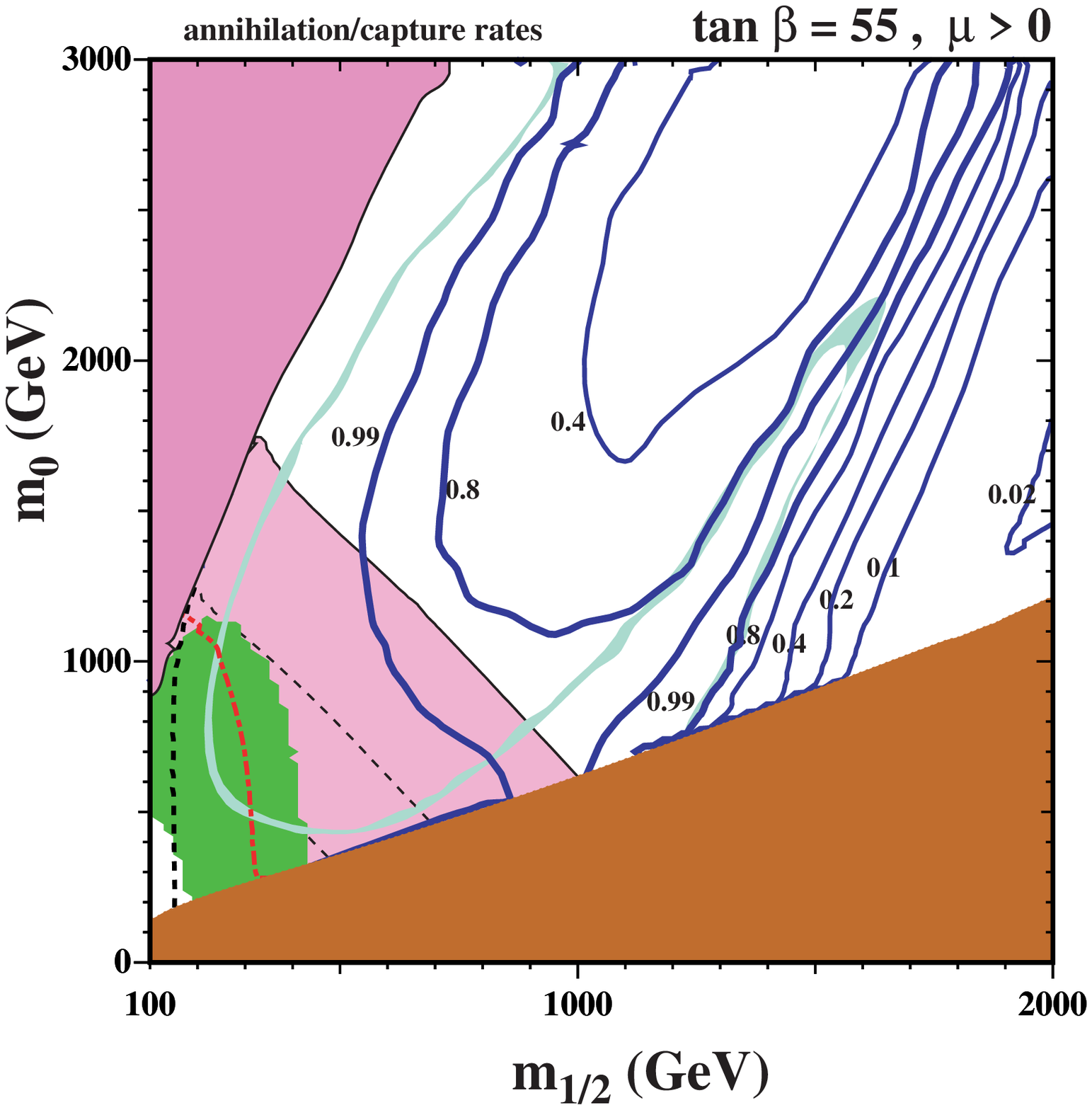}
  \caption{\it
    \coloronlinestatement
    The $(m_{1/2},m_0)$ planes for the CMSSM for $A_0 = 0$ and (left) 
    $\tanb = 10$, (right) $\tanb = 55$, showing contours of the
    ratio of solar dark matter annihilation and capture rates, as
    calculated using the AGSS09 model~\protect\cite{Serenelli:2009yc}
    and default values of the hadronic scattering matrix elements.
    Equilibrium corresponds to a ratio of unity, which is approached
    for small $m_{1/2}$ and large $m_0$.
    Also shown are the theoretical, phenomenological, experimental and
    cosmological constraints described in the text.
  }
  \label{fig:capann}
\end{figure*}

As already mentioned, it might be thought that the LSP density within
the Sun is in equilibrium, so that the annihilation rate is the same
as the capture rate.
However, as seen in \reffig{capann}, this is not in general true
within the class of CMSSM scenarios discussed here~\footnote{
  In calculating the contours in this figure,
  we use the numerical AGSS09~\cite{Serenelli:2009yc} solar model and
  the default values of the hadronic matrix elements 
  discussed below.}.
In \reffig{capann}, we show contours of the ratio of the annihilation
rate to the capture rate. Contours where this ratio exceeds 0.8 are
in bold for clarity.
In particular, for $\tanb = 10$, as seen in the left panel of
\reffig{capann}, the LSP annihilation rate is in general much smaller
than the capture rate, indicating that equilibrium is very poor
approximation. The only regions in which equilibrium is approached are
at very low $m_{1/2}$ where the LEP Higgs bound (red dash-dotted line)
is violated, and close to the WMAP strip in the focus-point region,
where annihilation exceeds 99\% of the capture rate. In contrast, the
annihilation rate falls to less than 1\% of the capture rate along the
coannihilation strip at large $m_{1/2}$.  
The loops centered at $m_{1/2} \approx 140$ GeV are due
to a drop in the relic density due to rapid s-channel annihilation
through the light Higgs boson.  As a consequence, there are drops in
both the capture and annihilation rates there.  The kink and loop in
the $(m_{1/2}, m_0)$ plane for $\tanb = 10$ (left panel of
\reffig{capann}) centered 
at $m_{1/2} \sim 200$~GeV, where the ratio of the annihilation and
capture rates is relatively high, is due to the enhancement of
annihilation just above the $W^+ W^-$ threshold.  The fluctuations in
the contours of the ratio of the annihilation and capture rates for
$m_{1/2} \sim 400$~GeV in the left panel of \reffig{capann} reflect an
increase in the annihilation rate above the ${\bar t} t$ threshold. 

For $\tanb = 55$, as shown in the right panel of \reffig{capann},
there are larger regions where the annihilation
rate exceeds 99\% of the capture rate, extending to
$m_{1/2} \sim 500$~GeV and more.  In particular, equilibrium is
approached in a significant fraction of the coannihilation strip, as
well as along the focus-point strip, and also in the neighbourhood of
the rapid-annihilation funnel that extends to $m_{1/2} \sim 1500$~GeV,
and beyond.  This reflects the rapid increase in the annihilation cross
section as one enters this funnel. We also note that the contours of
the ratio of the annihilation and capture rates vary rapidly in the
neighbourhood of the rapid-annihilation funnel, particularly at larger
$m_{1/2}$. As one approaches the funnel, the annihilation rate increases,
thus increasing the ratio.  At higher $m_{1/2}$, inside the funnel, the
relic density is suppressed and the rates are scaled.  At still larger
$m_{1/2}$ the elastic cross section is 
suppressed and equilibrium is not recovered.  

Being out of equilibrium has important consequences when we examine
how variations in solar models and, later, scattering cross sections
affect the annihilation rates and neutrino fluxes.  In regions where
capture and annihilation are in equilibrium ($t_{\odot} \gg \tau$),
$2 \Gamma \propto C$ so that, \eg\ a 10\% drop in the capture rate
would lead to a $\sim$10\% drop in the annihilation rate.  Since, for
a given CMSSM model, the neutrino flux is proportional to the
annihilation rate, the neutrino flux would likewise decrease by
$\sim$10\%.  However, when capture and annihilation are well out of
equilibrium ($t_{\odot} \ll \tau$), $2 \Gamma \propto C^2$.  In
this case, a 10\% drop in the capture rate would lead to a $\sim$19\%
drop in the annihilation rate and neutrino flux.  Thus, the non-linear
dependence of \refeqn{ar} on the capture rate can amplify
variations in the annihilation rate and hence the neutrino flux relative
to the variations in the capture rate found in the various cases
considered later in the paper.

\subsection{Dependence on the Solar Model}

As discussed above, the rate at which the LSP is captured by the Sun
involves determining both the rate at which galactic LSPs scatter in the
Sun and the fraction of such scatters in which the LSP loses enough
energy to become gravitationally bound to the Sun.  The former depends
on the LSP-nucleus scattering cross sections, while the latter depends
on the mass of the nucleus off of which the LSP scatters; both factors
may vary significantly among the different isotopes present in the Sun.
The capture rate is thus sensitive to the abundances of elements in
the Sun.

\begin{figure*}
  \insertdoublefig{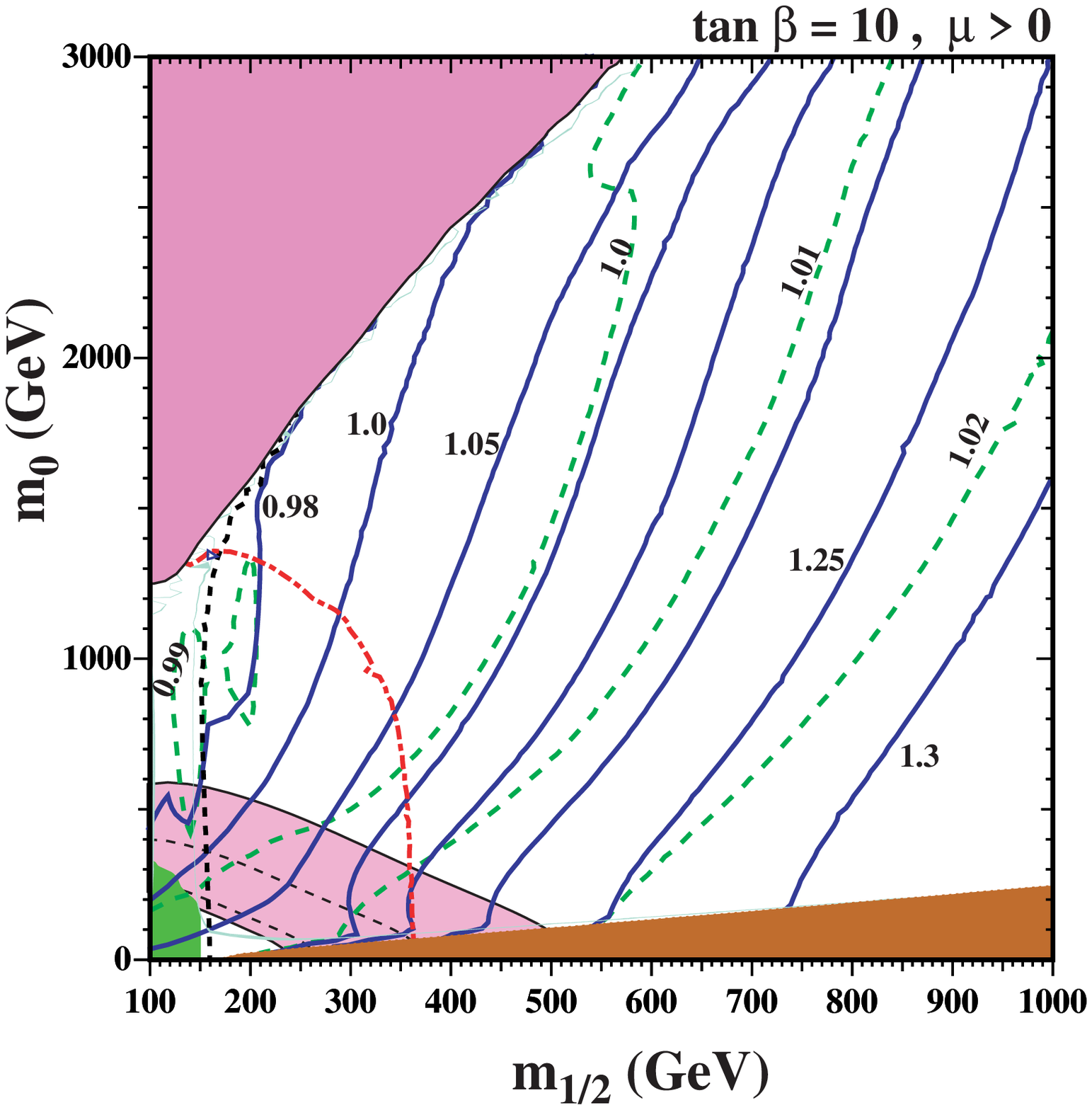}{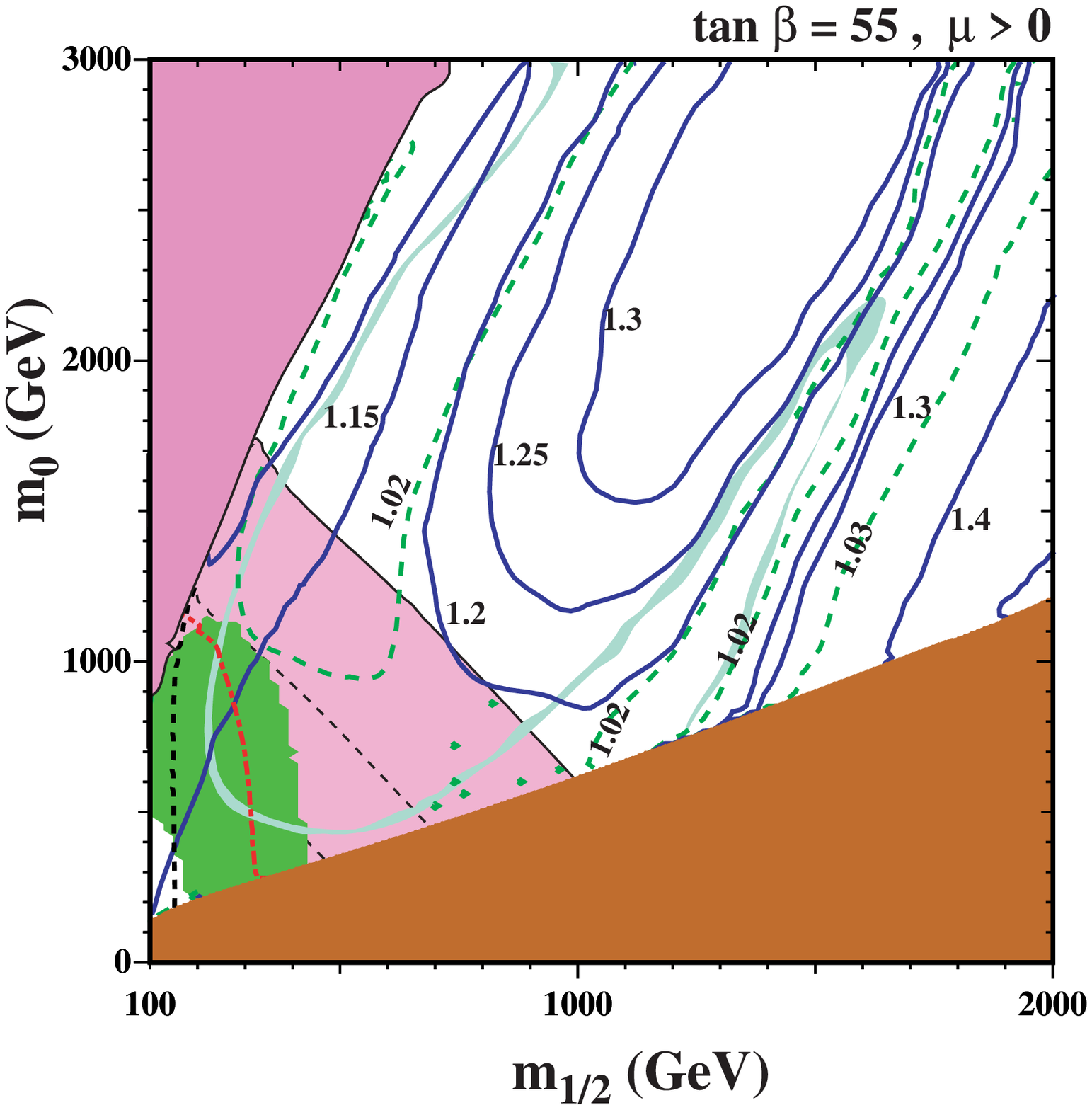}
  \caption{\it
    \coloronlinestatement
    The $(m_{1/2},m_0)$ planes for the CMSSM for $A_0 = 0$ and (left) 
    $\tanb = 10$, (right) $\tanb = 55$, showing contours of the
    ratios of the solar dark matter annihilation rates calculated using
    the GS98 (solid) and AGSS09ph (dot-dashed) solar models to the
    annihilation rates found using the fiducial AGSS09
    model~\protect\cite{Serenelli:2009yc}.
    Also shown are the theoretical, phenomenological, experimental and
    cosmological constraints described in the text.
  }
  \label{fig:solarmodels}
\end{figure*}

Serenelli \etal\ \cite{Serenelli:2009yc} have generated models of the
Sun for several proto-solar isotopic abundances\footnote{
  We include all 29 isotopes/elements maintained in these models when
  doing our calculations.
  }.  The models AGSS09
and AGSS09ph are based on the most recent estimates of these abundances
\cite{Asplund:2009fu}; the latter model differs from the former in that
photospheric measurements instead of meteoritic measurements are used
to estimate abundances of refractory elements.
Solar models based on recent abundance estimates are in conflict with
helioseismological measurements of the Sun; Serenelli \etal\ provide
an additional model (GS98), based on older abundance estimates
\cite{Grevesse:1998bj}, that is compatible with helioseismology.
We show in \reffig{solarmodels} comparisons of the
annihilation rates in the AGSS09ph and GS98 models with the
fiducial case (AGSS09) in the planes for $\tanb = 10$ and $55$.

The primary difference between the abundance sets and their
corresponding solar models is an ${\cal O}(10)$\% change in the
metallicity in
the Sun.  Whilst these metals represent less than 2\% of the Sun's mass,
we show later that capture via spin-independent interactions can
be significantly enhanced by heavy nuclei due to a rapid increase in
the scattering cross section.  Moreover, the capture of heavy LSPs is
favored kinematically by scattering on heavier elements.  Due to these
effects, these trace heavy elements can, in some cases, contribute a
sizeable portion of the total capture rate.  We see in
\reffig{solarmodels} that there is at most a 4\% difference in the
annihilation rates between the models based upon the two most recent
abundance estimates (AGSS09 and AGSS09ph), so that the discrepancy
between meteoritic and photospheric measurements is not a significant
issue in estimating annihilation rates in CMSSM models.  However,
the model based on older abundance estimates that more closely matches
helioseismology measurements (GS98) predicts annihilation rates as
much as 40\% higher.  In both cases, the discrepancy grows larger as
$m_{1/2}$ increases, since SI scattering, which occurs mainly off of the
trace heavy elements that differ among the models, becomes more
important due to the aforementioned effects.
In addition, part of the larger variation at higher $m_{1/2}$ can be
attributed to non-equilibrium between capture and annihilation as
discussed previously.  For regions noted in \reffig{capann} as being
out of equilibrium, the variation in the capture rate for the different
solar models is roughly half that of the variation in the annihilation
rate.  For regions in equilibrium, the variation in the capture rate
is roughly the same as the variation in the annihilation rate.

It is unclear whether the discrepancies between the recent models and
helioseismology is due to inaccuracies in the solar modelling or
due to the most recent abundance estimates being incorrect; 
see~\cite{Serenelli:2009ww} and references therein for
further discussion of this issue.
For this reason, uncertainties in the solar model still limit the
precision to which annihilation rates (and, thus, neutrino fluxes) can
be predicted for a given CMSSM model, but are not important for our
conclusions.

\subsection{Approximating the Solar Potential}

As shown by Gould~\cite{Gould:1992xx}, the calculation of the capture
rate can be simplified if the solar potential is approximated by
\begin{equation} \label{eqn:PhiApprox}
  \Phi(r) = \Phi_c - \frac{M(r)}{M_{\odot}} (\Phi_c - \Phi_s) \, ,
\end{equation}
where $M(r)$ is the mass of the Sun interior to a radial distance $r$,
$\Phi_c$ and $\Phi_s$ are the gravitational potentials at the center
and surface of the Sun, respectively, and $\Phi(r)$ is the potential
at an arbitrary point in the Sun.  This approximation yields an
analytical form for the full capture rate, instead of requiring a
numerical integration over the solar profile as necessary in the exact
case.  The analytical form is lengthy and can
be found in Ref.~\cite{Gould:1992xx}.

The accuracy of the approximation is improved by choosing values for
$\Phi_c$ and $\Phi_s$ such that \refeqn{PhiApprox} more closely
follows the true potential profile (as determined in the numerical
models) rather than using the actual central and surface potentials.
We use $v_c$ = 1355~km/s and $v_s$ = 818~km/s for the central and
surface escape velocities (where $\Phi(r) = -\frac{1}{2} v^2(r)$),
which is obtained from a linear fit to the inner 90\% (by mass) of
the Sun in the AGSS09 model.  The outer 10\% is ignored in the fit
as the potential can only be well approximated by a line for the
inner $\sim$90\%; inclusion of the outer layer makes the fit much
poorer in the rest of the Sun.  In addition, the dependence of
\refeqn{dCidM} on the potential generally becomes most significant
when capture is suppressed at the surface relative to the center;
thus, a better fit to the inner regions of the Sun is preferred to
a fit to the entire solar profile.  For the parameters given above,
\refeqn{PhiApprox} varies by less than 5\% from the actual potential
in the AGSS09 model for the inner 90\% of the Sun.  The fits to
$v_c$ and $v_s$ for other solar models do not change significantly
from the above values.

The Gould approximation requires the further assumption that the
abundances of elements are uniform throughout the star.  This is a
reasonable approximation for elements heavier than Oxygen, but is
not a good approximation for lighter elements.  Hydrogen and Helium,
in particular, are much more and less abundant, respectively, in the
center of the Sun than elsewhere due to the hydrogen fusion
occurring there.
Carbon, Nitrogen, and Oxygen abundances also vary throughout the star,
under the influence of the CNO
cycle. However, the combined CNO abundance is fairly uniform and,
since these three elements have similar masses and cross sections,
there is little loss in accuracy in taking a uniform abundance for
them.

\begin{figure*}
  \insertdoublefig{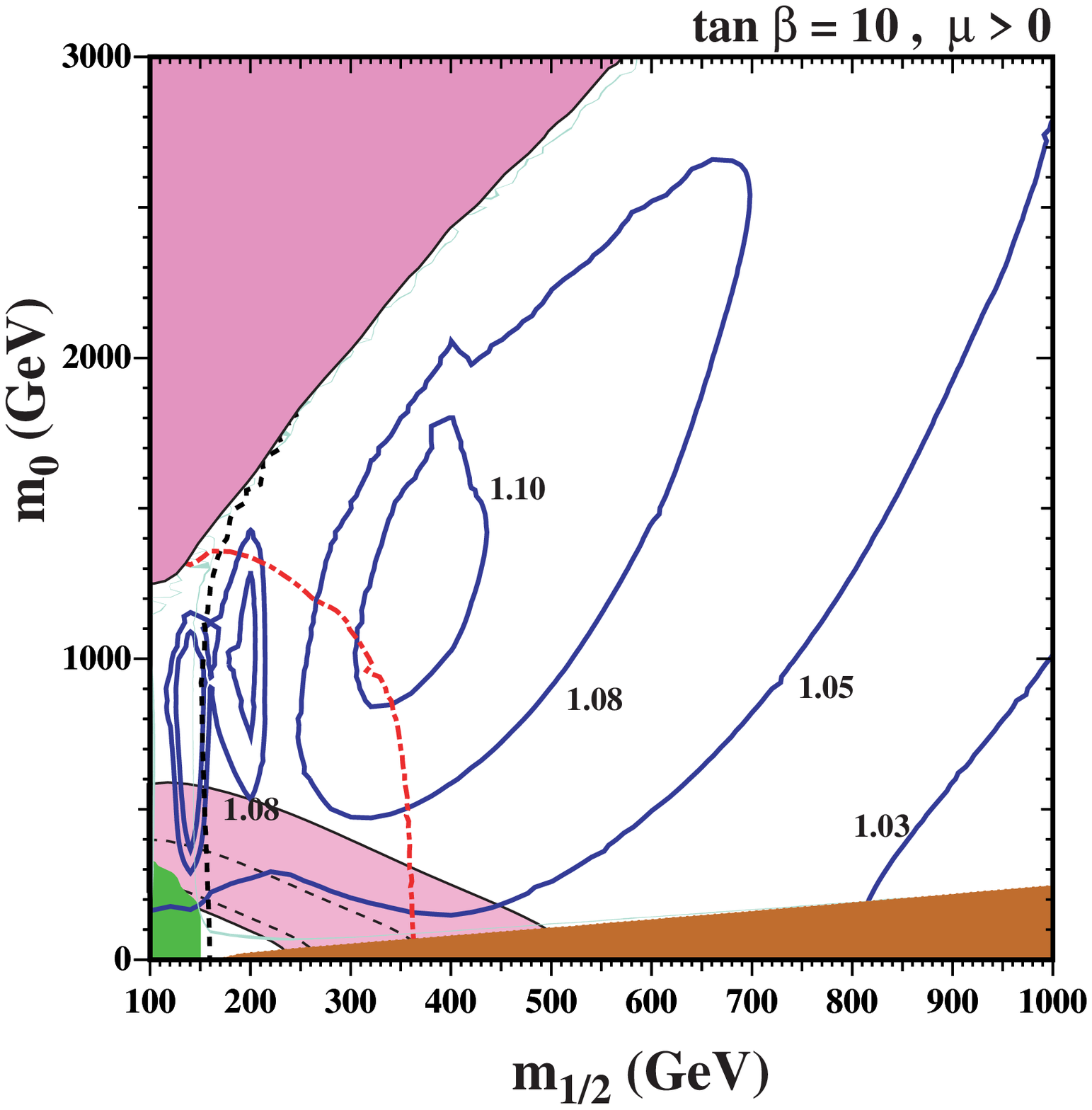}{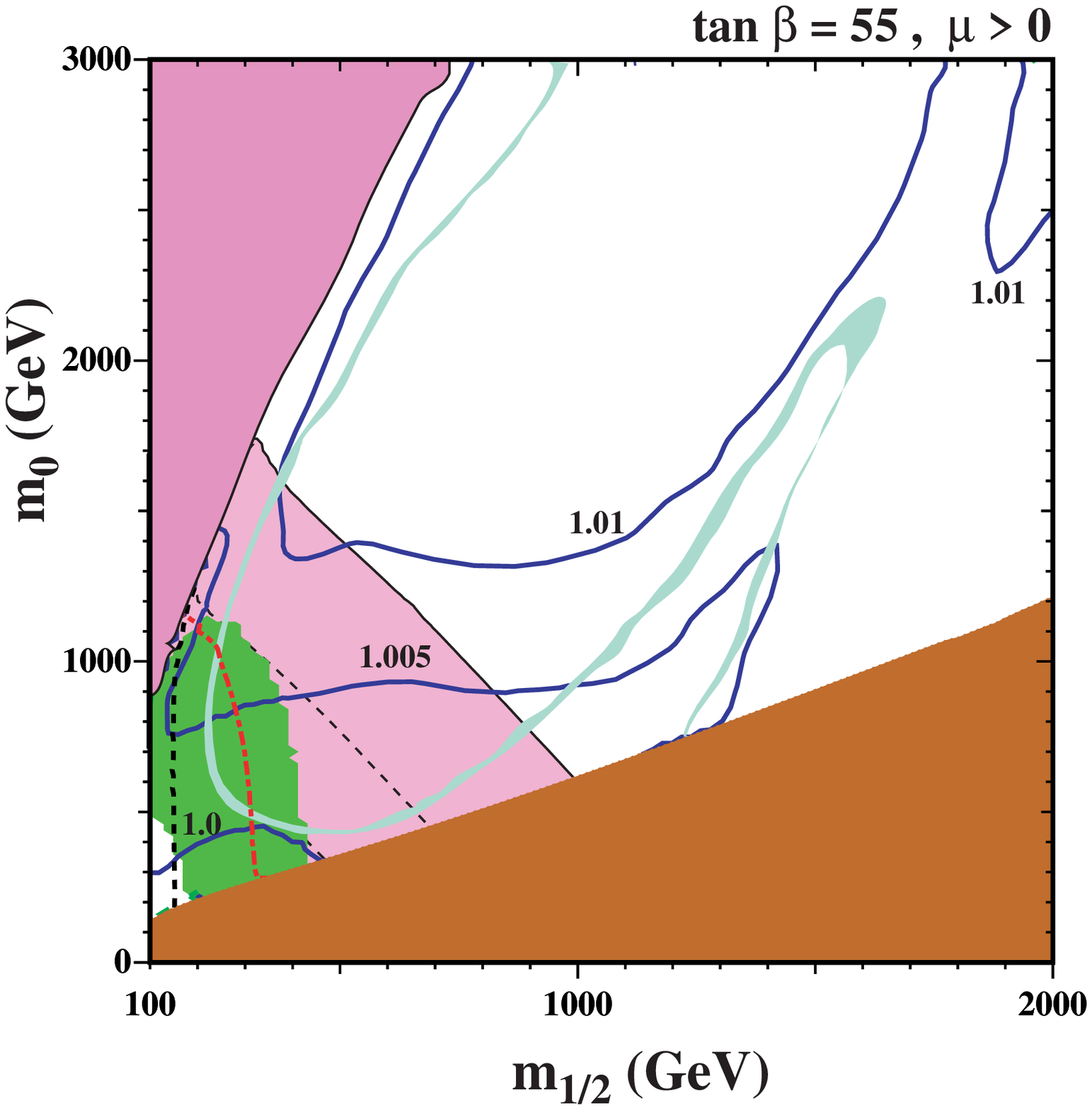}
  \caption{\it
    \coloronlinestatement
    The $(m_{1/2},m_0)$ planes for the CMSSM for $A_0 = 0$ and (left) 
    $\tanb = 10$, (right) $\tanb = 55$, showing contours of the
    ratio of the dark matter annihilation rates calculated using a
    full numerical integration over the AGSS09 solar
    profile~\protect\cite{Serenelli:2009yc}
    and using an analytical approximation as proposed by Gould.
    Also shown are the theoretical, phenomenological, experimental and
    cosmological constraints described in the text.
    }
  \label{fig:gould}
\end{figure*}

Looking at \reffig{gould}, we see that the numerical integration and
the Gould approximation agree quite well in general, always
within 11\% in the $\tanb = 10$ plane, and better at high
$m_{1/2}$ and in the focus-point region.  For $\tanb = 55$, the
accuracy is even better, with the approximation never varying more
than 2\% from the numerical result.  The different levels of
agreement between the numerical and analytical estimates have two
sources: (1) varying levels of contribution between the spin-dependent
and -independent scattering, and (2) amplification of differences
when annihation and capture are out of equilibrium.

The capture rate itself is nearly uniformly overestimated
by $\sim$8\% for spin-dependent interactions and underestimated by
$\sim$0.4\% for spin-independent interactions when the Gould
approximation is used.  The approximation to
the potential, \refeqn{PhiApprox}, leads to a $\lae$1\% underestimate.
The assumed uniformity of Hydrogen throughout the Sun, however,
gives a $\sim$8\% overestimate of the capture via scattering off of
Hydrogen.
As Hydrogen accounts for over 99\% of the captures via
spin-dependent scatters~\footnote{
  We have included spin-dependent scattering on all significant
  isotopes in the Sun (notably Helium-3 and Nitrogen-14), but find
  that elements other than Hydrogen account for much less than 1\%
  of the capture from this type of scattering in the models considered
  here.
  }, but only a negligible portion of the capture via spin-independent
scatters, spin-dependent capture is thus overestimated by $\sim$8\%,
while spin-independent capture remains underestimated by $\sim$0.4\%.
Note that Helium is not the dominant nuclear target for spin-independent
capture, so the poor assumption of uniform Helium abundance has little
bearing on the results.
As seen in \reffig{sds} and discussed in more detail below,
spin-dependent scattering is a significant (and dominant for
$\tanb = 10$) contribution to the total capture rate at small
$m_{1/2}$, but is generally less significant at higher $m_{1/2}$,
where spin-independent capture takes over.  Thus, the Gould
approximation becomes a better approximation at higher $m_{1/2}$.

The total capture rate varies by up to 6\% with the approximation,
with the largest difference occurring for $\tanb = 10$, where
spin-dependent scattering is more significant.
Again, non-equilibrium amplifies the variations, increasing 1-6\%
variations in the capture rate to as much as 2-11\% variations in the
annihilation rate.
The largest variation in the annihilation rate, which occurs for
$m_{1/2} \sim 300-500$~GeV in the $\tanb = 10$ plane (left panel of
\reffig{gould}) occurs in the region where spin-dependent capture
dominates and simultaneously annihilation is out of equilibrium.  For
higher values of $m_{1/2}$ in the $\tanb = 10$ plane, annihilation is
still out of equilibrium, but spin-dependent scattering becomes less
important.  Though the non-equilibrium amplifies variations in the
capture rate here, the capture rate for spin-independent scattering is
much more accurate than the spin-dependent case, so the overall
annihilation rate in the approximation becomes closer to the true
value.
For the $\tanb = 55$ case (right panel of \reffig{gould}),
spin-independent scattering dominates everywhere.  Because the
spin-independent capture rate is very accurately approximated, the
approximation to the annihilation rate is quite accurate for
$\tanb = 55$.  The shapes of the contours in the right panel of
\reffig{gould} mainly result from the changing levels of equilibrium
between annihilation and capture rather than from changes in the
relative importance of spin-dependent and spin-independent scattering.
In the focus-point and funnel regions, where annihilation and capture
are in equilibrium, the variations in the annihilation rate reflect
those of the capture rate ($\sim$0.5-1\%).  In the area between
the focus point and funnel regions, where equilibrium is not
maintained, the variation in the annihilation rate grows to roughly
twice that of the capture rate ($\sim$1-2\%).

In principle, numerical integration over the true solar profile is
certainly preferable to using the Gould approximation, because of its
greater accuracy.  However, there may be instances, such as
computationally-intensive scans over MSSM parameter space, where
avoiding a numerical integration would be useful~\footnote{The same
  could be accomplished in some cases---but not all---by
  tabulating the integration, as done in DarkSUSY~\cite{darksusy}.}.
For example, using alternative form factors or LSP velocity
distributions may require the integrals in \refeqn{dCidM} to be
evaluated numerically, which could make determining the capture rate
a computationally-slow process where reducing the number of numerical
integrations would be beneficial time-wise.
Also, we note that using this approximation introduces inaccuracies
no worse than those due to choices in solar models and uncertainties in
hadronic parameters, as discussed below.
There is also room for improvement in this approximation as the chief
source of its inaccuracy---the assumed uniformity of
the Hydrogen abundance---could be significantly alleviated
by, \eg, breaking \refeqn{PhiApprox} into two parts, corresponding to
the inner and outer portions of the Sun, and using more appropriate
Hydrogen abundances in each part.

\section{\label{sec:hadronic} Standard Model Particle Physics
                              Uncertainties}

We now turn to the uncertainties in the annihilation rates induced
by the uncertainties in the hadronic matrix elements that enter into
dark matter scattering rates, which are listed in
\reftab{params}. They include those in the quark masses,
expressed as $m_{\rm{d,c,b,t}}$ and the ratios $\mumd$ and $\msmd$,
those in the matrix elements $\langle N | {\bar q} q | N \rangle$ --
which are related to the change in the nucleon mass
due to non-zero quark masses, denoted by $\sigma_0$, and 
therefore to the $\pi$-nucleon
$\sigma$ term $\SigmapiN$ as discussed later -- and the
axial-current matrix elements $\langle N | {\bar q} \gamma_\mu
\gamma_5 q | N \rangle$, which are related to the quantities
$\Deltaps, \athree$ and $\aeight$, as also discussed later. 
The uncertainties in the elastic scattering cross section induced
by the uncertainties in the quark masses, apart from the top quark,
are negligible. However, cross-section uncertainties
induced by the uncertainties in the matrix elements 
$\langle N | {\bar q} \gamma_\mu \gamma_5 q | N \rangle$ and
$\langle N | {\bar q} q | N \rangle$ are important, as we discuss below.
In particular, the uncertainties induced by our ignorance of $\Deltaps$
and particularly $\SigmapiN$ should not be neglected.

\begin{table}
  \addtolength{\tabcolsep}{1em}
  \begin{tabular}{lc@{$\,\pm\,$}cc}
    \hline \hline
    \mumd & 0.553 & 0.043    &~\cite{Leutwyler:1996qg} \\
    \md   & 5     & 2 MeV    &~\cite{Amsler:2008zzb} \\
    \msmd & 18.9  & 0.8      &~\cite{Leutwyler:1996qg} \\
    \mc   & 1.27  & 0.09 GeV &~\cite{Amsler:2008zzb} \\
    \mb   & 4.25  & 0.15 GeV &~\cite{Amsler:2008zzb} \\
    \mt   & 173.1 & 1.3 GeV  &~\cite{:2009ec} \\
    \hline
    $\sigma_0$ & 36 & 7 MeV &~\cite{Borasoy:1996bx} \\
    \SigmapiN  & 64 & 8 MeV &~\cite{Pavan:2001wz,Ellis:2005mb} \\
    \hline
    \athree  & 1.2694 & 0.0028 &~\cite{Amsler:2008zzb} \\
    \aeight  & 0.585  & 0.025  &~\cite{Goto:1999by,Leader:2002ni} \\
    \Deltaps & -0.09  & 0.03   &~\cite{Alekseev:2007vi} \\
    \hline \hline
  \end{tabular}
  \caption[Parameters]{\it
    Strong-interaction parameters used to determine neutralino-nucleon
    scattering cross sections, with estimates of their experimental
    uncertainties. As discussed in the text,
    the most important uncertainties are those in $\sigma_0, \SigmapiN$
    and $\Deltaps$. We use as defaults for them the central values
    given in the Table.
    }
  \label{tab:params}
\end{table}

\begin{figure*}
  \insertdoublefig{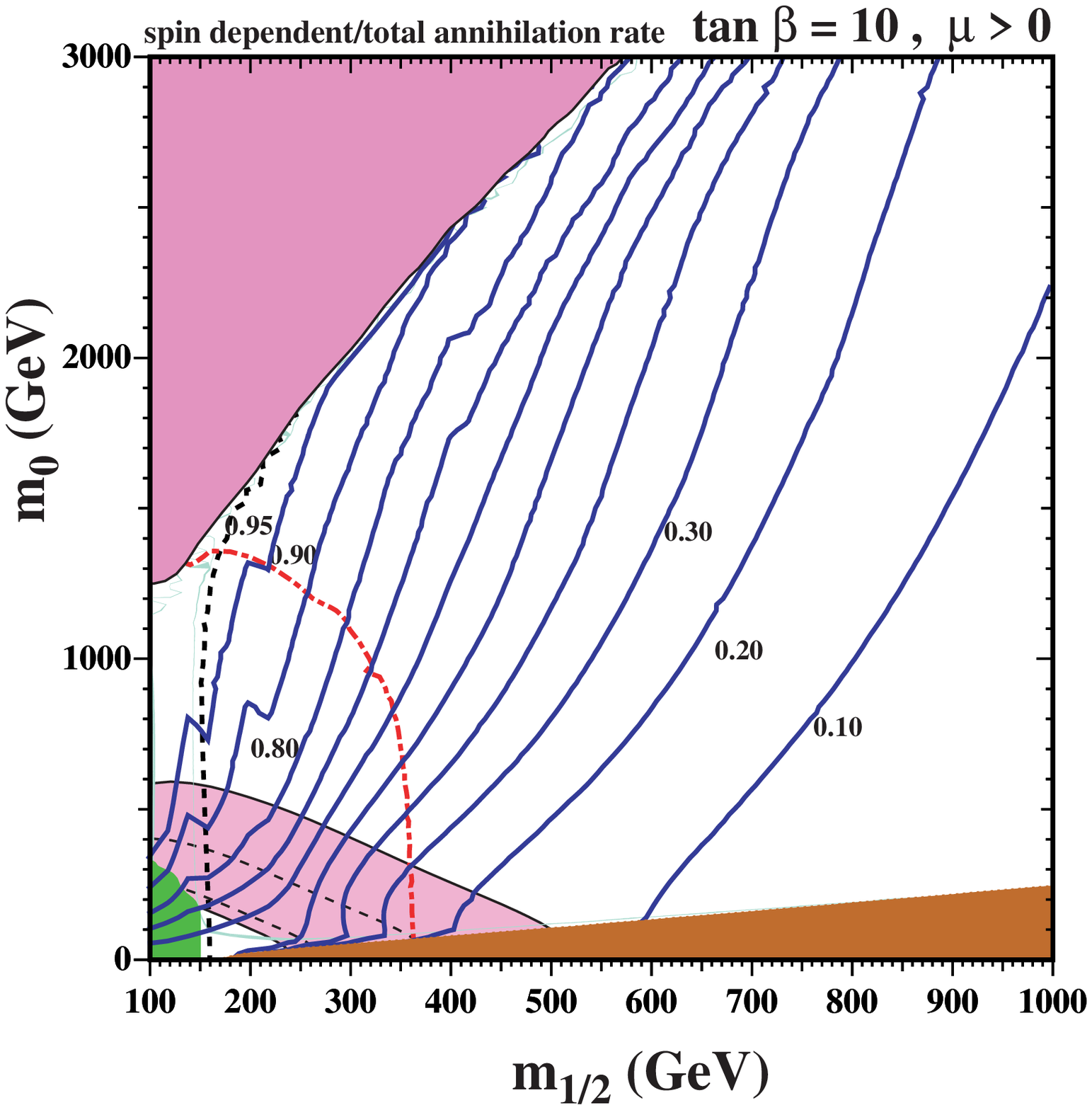}{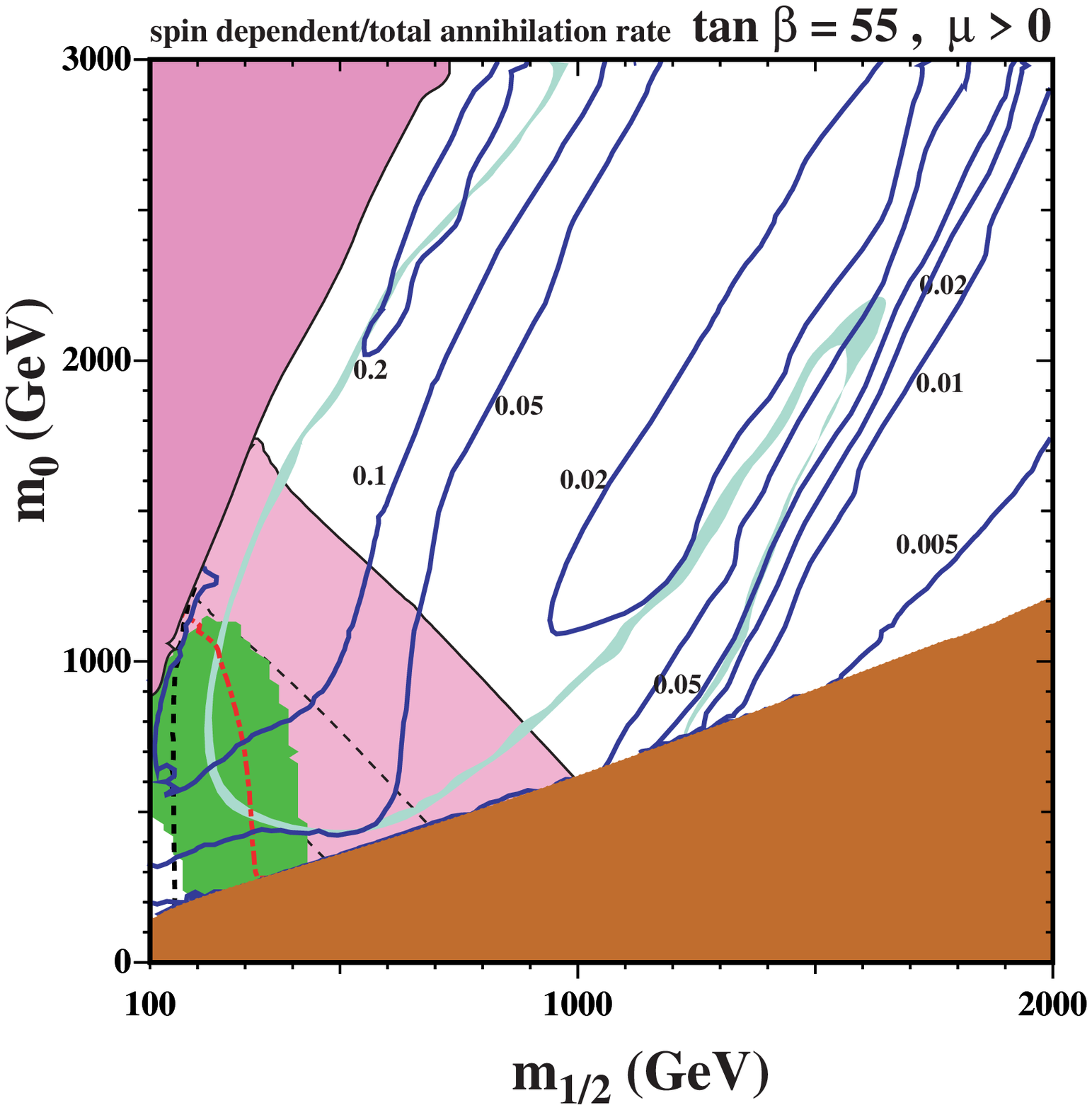}
  \caption{\it
    \coloronlinestatement
    The $(m_{1/2},m_0)$ planes for the CMSSM for $A_0 = 0$ and (left) 
    $\tanb = 10$, (right) $\tanb = 55$, showing contours of the
    ratio of the solar dark matter annihilation rate calculated
    using only spin-dependent scattering to the total annihilation
    rate including also spin-independent scattering.
    Also shown are the theoretical, phenomenological, experimental and
    cosmological constraints described in the text.
    }
  \label{fig:sds}
\end{figure*}

As already mentioned, it might be thought that the dominant dark-matter
scattering process inside the Sun is spin-dependent scattering on the
protons that comprise most of the solar mass. However, this is not
necessarily the case in the CMSSM, as seen in \reffig{sds} which
shows
the annihilation rate when only spin-dependent scattering is
included, relative to the total annihilation rate when both
spin-dependent and -independent scattering is included.
We see in the left panel for $\tan \beta = 10$  that, 
whereas spin-dependent scattering is indeed
dominant at small $m_{1/2}$ and large $m_0$, this is a poor
approximation already in the portion of the coannihilation strip that
is favoured by $g_\mu - 2$, and spin-independent scattering actually
\textit{dominates} at large $m_{1/2}$ along the coannihilation strip.
For $\tanb = 55$ (right panel), spin-independent scattering dominates
across the plane and signicantly so in the coannihilation strip and in
the rapid-annihilation funnel~\footnote{
  In calculating the contours in this figure, we use the
  default values of $\Deltaps$ and $\SigmapiN$ discussed below.
  }.
The increase in the relative importance of spin-independent scattering
at large $m_{1/2}$ may be traced to the importance of the reduced
neutralino mass $m_r$ in the kinematics of the scattering process. As
the neutralino mass increases, the dependence on $m_r$ disfavours
capture via scattering off light nuclei such as the proton that
dominates spin-dependent scattering, and increases the relative
importance of spin-independent scattering, which is dominated by
scattering off more massive nuclei. Furthermore, because the
spin-independent cross section increases with $\tanb$, its importance
is amplified in the right panel for $\tan \beta = 55$. As in previous
figures, we see in
the right panel of \reffig{sds} for $\tanb = 55$ that the contours of
the ratio are relatively close together in the neighbourhood of the
rapid-annihilation funnel. In this region the annihilation rate varies
rapidly and approaches equilibrium with the capture rate.
Without spin-independent scattering, the capture rate there would be
reduced, driving annihilation out of equilibrium and suppressing the
annihilation rate and the resulting neutrino flux.

The results shown in \reffig{sds} imply that one must discuss the
uncertainties in both spin-dependent and -independent LSP scattering,
as we discuss in  more detail in the subsequent subsections.

\subsection{Sensitivity to Spin-Dependent Scattering}

The spin-dependent (SD) part of the elastic $\chi$-nucleus cross
section can be written as
\begin{equation} \label{eqn:sigmaSD}
  \sigma_{\rm SD} = \frac{32}{\pi} G_{F}^{2} m_{r}^{2}
                    \Lambda^{2} J(J + 1) \; ,
\end{equation}
where $m_{r}$ is the reduced neutralino mass, $J$ is the spin 
of the nucleus,
\begin{equation} \label{eqn:Lambda}
  \Lambda \equiv \frac{1}{J} \left(
                 a_{\rmpp} \langle S_{\rmpp} \rangle
                 + a_{\rmnn} \langle S_{\rmnn} \rangle
                 \right) \; ,
\end{equation}
$\langle S_{\rmpp} \rangle$ and $\langle S_{\rmpp} \rangle$ are the
spin content of the proton and neutron groups, respectively
\cite{Bednyakov:2004xq}, and
\begin{equation} \label{eqn:aN}
  a_{\rmpp} = \sum_{q} \frac{\alpha_{2q}}{\sqrt{2} G_{f}}
              \Deltapq{q} , \qquad
  a_{\rmnn} = \sum_{i} \frac{\alpha_{2q}}{\sqrt{2} G_{f}}
              \Deltanq{q} \; .
\end{equation}
The factors $\DeltaNq{q}$ parametrize the quark spin content of the
nucleon and are significant only for the light (u,d,s) quarks.
A combination of experimental and theoretical results tightly
constrain the linear combinations~\cite{Amsler:2008zzb}
\begin{equation} \label{eqn:a3}
  \athree \equiv \Deltapq{\rmu} - \Deltapq{\rmd}
          = 1.2694 \pm 0.0028
\end{equation}
and~\cite{Goto:1999by,Leader:2002ni}
\begin{equation} \label{eqn:a8}
  \aeight \equiv \Deltapq{\rmu} + \Deltapq{\rmd} - 2 \Deltapq{\rms}
          = 0.585 \pm 0.025 .
\end{equation}
However, determination of the individual $\DeltaNq{q}$ requires a third
piece of information, usually taken to be the strange spin
contribution $\Deltapq{\rms}$, as extracted from inclusive
deep-inelastic lepton-nucleon scattering.
Using, \eg, the recent COMPASS result~\cite{Alekseev:2007vi}, one has
\begin{eqnarray} \label{eqn:Deltaps}
  \Deltapq{\rms} &=& -0.09 \pm 0.01 \, \text{(stat.)} \,
                         \pm 0.02 \, \text{(syst.)}
  \ifmulticol{\nonumber \\ &\approx&}{\approx}
                     -0.09 \pm 0.03 \; ,
\end{eqnarray}
where, conservatively, we have combined linearly the statistical and
systematic uncertainties. Using this range for $\Deltapq{\rms}$, 
we may express $\DeltaNq{\rmu,\rmd}$ as follows in terms
of known quantities:
\begin{alignat}{3}
  \label{eqn:Deltapu}
  \Deltapq{\rmu}
    & \ = \ & \frac{1}{2} \left( \aeight + \athree \right) + \Deltaps
    & \ = \ & 0.84 \pm 0.03
    \; , \\
  \label{eqn:Deltapd}
  \Deltapq{\rmd}
    & \ = \ & \frac{1}{2} \left( \aeight - \athree \right) + \Deltaps
    & \ = \ & -0.43 \pm 0.03
    \; .
\end{alignat}
The above two uncertainties are almost completely correlated
with that of $\Deltaps$, however, and the uncertainties in $\athree$ 
and $\aeight$ are negligible by comparison.
We use the central value in \refeqn{Deltaps} as our default,
namely $\Deltapq{\rms} = -0.09$~\footnote{
  We recall that the proton and neutron scalar matrix elements are
  related by an interchange of $\Delta_{\rmu}$ and $\Delta_{\rmd}$,
  \ie, $\Deltanq{\rmu} = \Deltapq{\rmd}$,
  $\Deltanq{\rmd} = \Deltapq{\rmu}$ and
  $\Deltanq{\rms} = \Deltapq{\rms}$.
  }.

\begin{figure*}
  \insertdoublefig{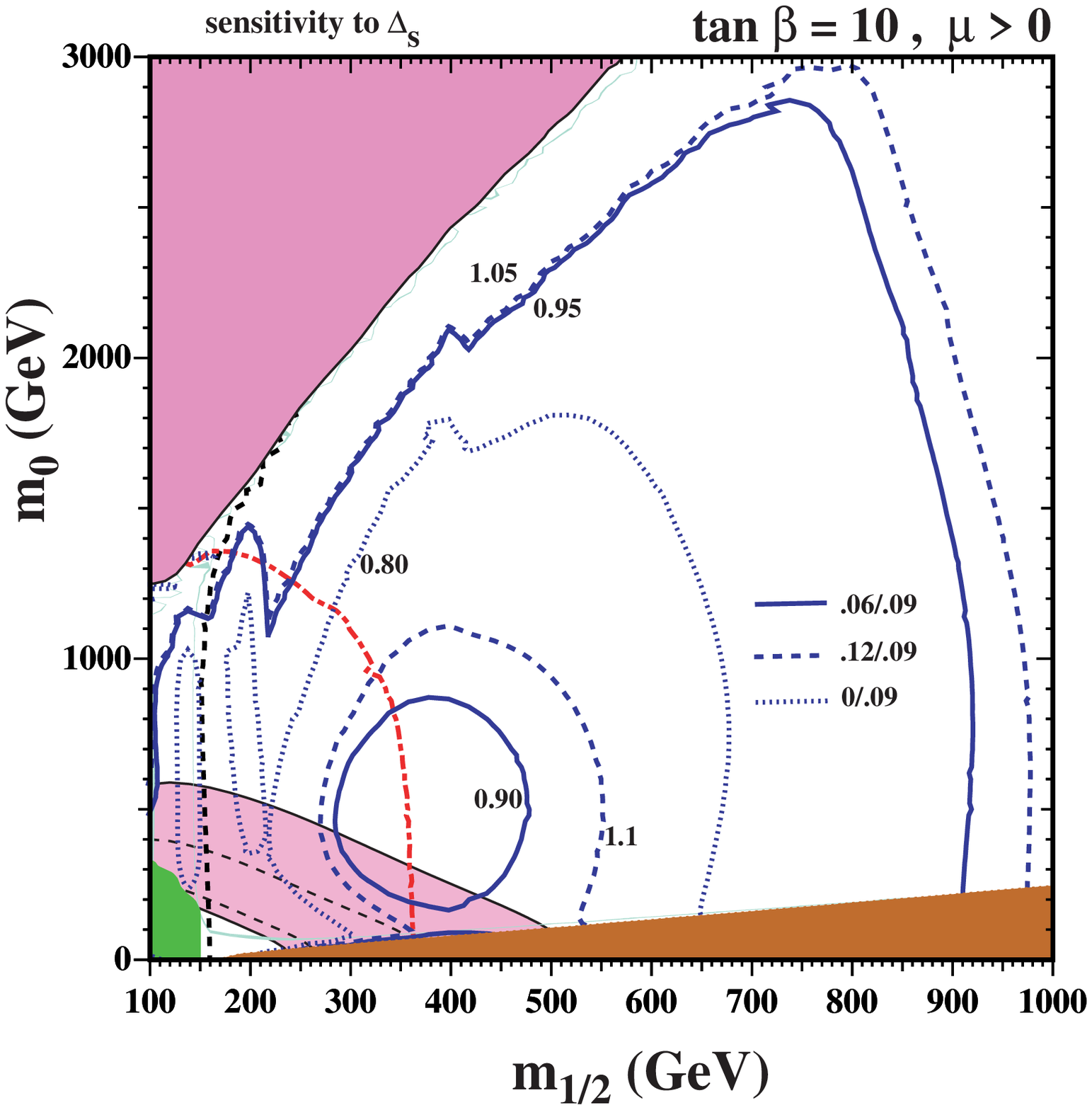}{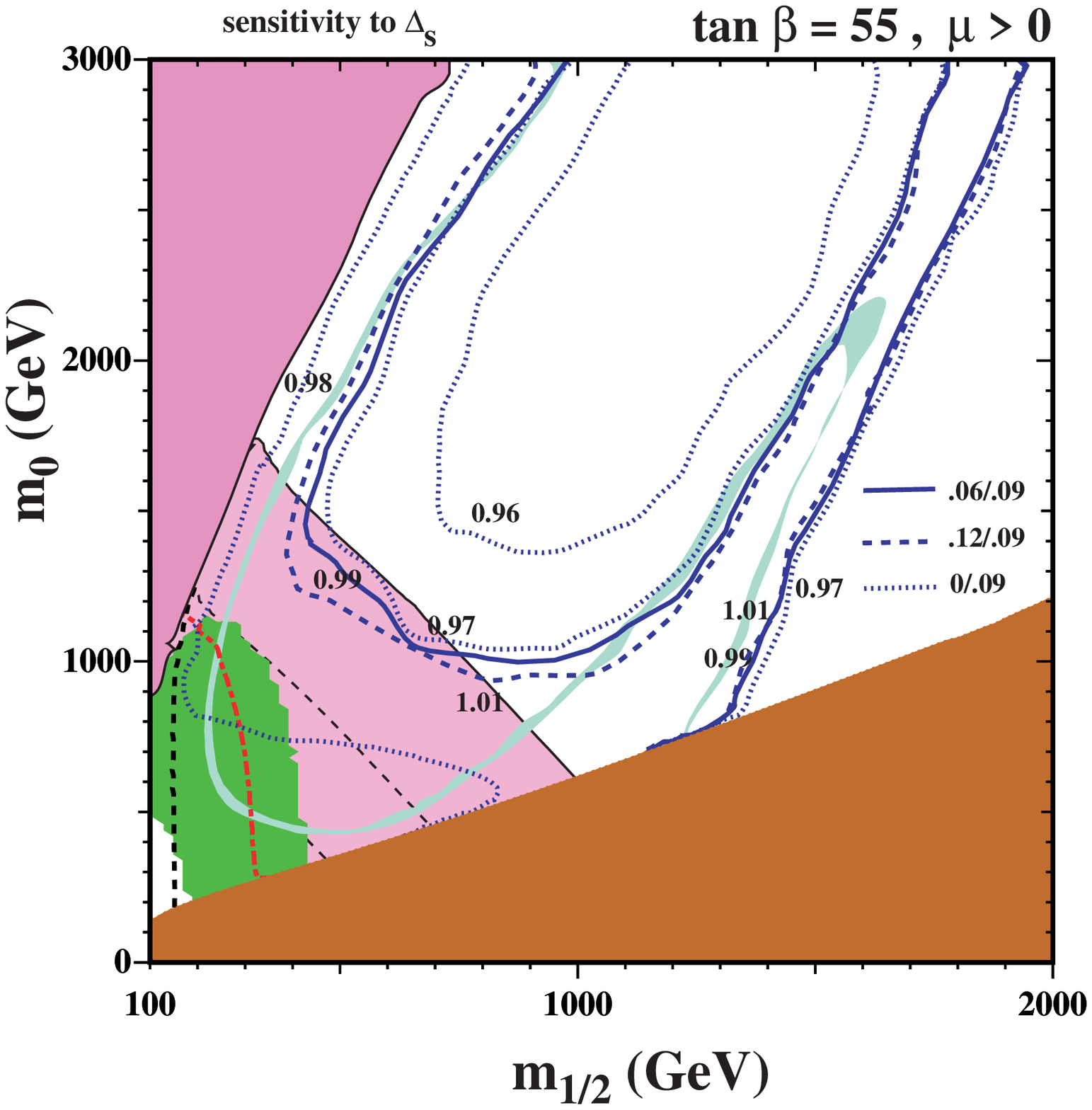}
  \caption{\it
    \coloronlinestatement
    The $(m_{1/2},m_0)$ planes for the CMSSM for $A_0 = 0$ and (left) 
    $\tanb = 10$, (right) $\tanb = 55$, showing contours of the
    ratios of the annihilation rates calculated assuming
    $\Deltaps = 0.00$ (dotted lines), $-0.06$ (solid lines) and
    $-0.12$ (dashed lines) to calculations with the default value
    $\Deltaps = -0.09$.
    Also shown are the theoretical, phenomenological, experimental and
    cosmological constraints described in the text.
    }
  \label{fig:dss}
\end{figure*}

For comparison, in \reffig{dss} we compare the
annihilation rates calculated using this default value and with
$\Deltapq{\rms} = -0.12$ and $-0.06$. We recall, however, that analyses
of $\pi$ and K meson production in
deep-inelastic scattering have been interpreted as suggesting that
$\Deltapq{\rms}$ may be compatible with zero~\cite{Airapetian:2008qf}.
Accordingly, we also present in \reffig{dss} results for this extreme
value of $\Deltapq{\rms}$.

In general, the spin-dependent scattering rate increases and
decreases with $\Deltaps$, and the effect is quite symmetric for 
$\Deltaps \in (-0.12, -0.06)$. However, the effect is quite small for
$\Deltaps$ within this range, differing from the rate calculated
with our default assumption $\Deltaps = -0.09$ by at most $\sim 10$\%
for small $m_{1/2}$ when $\tanb = 10$ (left panel of \reffig{dss}) 
and considerably less for $\tanb = 55$ (right panel of 
\reffig{dss})~\footnote{
  The contours near the rapid-annihilation funnel again reflect the
  equilibrium/non-equilibrium effects on the annihilation rate
  discussed earlier.
  }.
However, the reduction may be considerably larger if $\Deltaps = 0$,
potentially exceeding 20\% at small $m_{1/2}$ when $\tanb = 10$.

\subsection{Sensitivity to Spin-Independent Scattering}

The spin-independent (SI) part of the cross section for scattering on
a nucleus $(Z, A)$ can be
written as
\begin{equation} \label{eqn:sigmaSI}
  \sigma_{\rm SI} = \frac{4 m_{r}^{2}}{\pi}
                    \left[ Z f_{p} + (A-Z) f_{n}  \right]^{2},
\end{equation}
where $m_r$ is the $\chi$-nuclear reduced mass and
\begin{equation} \label{eqn:fN}
  \frac{f_N}{m_N}
    = \sum_{q=\rmu,\rmd,\rms} \fNTq{q} \frac{\alpha_{3q}}{m_{q}}
      + \frac{2}{27} f_{TG}^{(N)}
        \sum_{q=\rmc,\rmb,\rmt} \frac{\alpha_{3q}}{m_q}
\end{equation}
for $N$ = p or n.  The parameters $\fNTq{q}$ are defined by
\begin{equation} \label{eqn:Bq}
  m_N \fNTq{q}
  \equiv \langle N | m_{q} \bar{q} q | N \rangle
  \equiv m_q \BNq{q} ,
\end{equation}
where~\cite{Shifman:1978zn,Vainshtein:1980ea}
\begin{equation} \label{eqn:fTG}
  f_{TG}^{(N)} = 1 - \sum_{q=\rmu,\rmd,\rms} \fNTq{q} .
\end{equation}
We take the ratios of the light quark masses from
\cite{Leutwyler:1996qg}:
\begin{equation} \label{eqn:mqmd}
  \frac{\mup}{\md} = 0.553 \pm 0.043 , \qquad
  \frac{\ms}{\md}  = 18.9 \pm 0.8 ,
\end{equation}
and the other quark masses are taken from~\cite{Amsler:2008zzb}, except
for the top mass, which is taken from the combined CDF and D0
result~\cite{:2009ec}.  These masses, as well as other experimental
quantities that will arise in the calculation of the hadronic matrix
elements, appear in \reftab{params}.

Following~\cite{Cheng:1988im}, we introduce the quantity:
\begin{equation} \label{eqn:z}
  z \equiv \frac{\Bpq{\rmu} - \Bpq{\rms}}{\Bpq{\rmd} - \Bpq{\rms}}
    = 1.49 ,
\end{equation}
which has an experimental error that is negligible compared with others
discussed below, and the strange scalar density
\begin{equation} \label{eqn:y}
  y \equiv \frac{2 \BNq{\rms}}{\BNq{\rmu} + \BNq{\rmd}}.
\end{equation}
In terms of these quantities, one may write
\begin{equation} \label{eqn:BdBu}
  \frac{\Bpq{\rmd}}{\Bpq{\rmu}}
   = \frac{2 + ((z - 1) \times y)}{2 \times z - ((z - 1) \times y)} \; .
\end{equation}
Proton and neutron scalar matrix elements are related by an interchange
of $B_{\rmu}$ and $B_{\rmd}$, \ie,
\begin{equation} \label{eqn:Bn}
  \Bnq{\rmu} = \Bpq{\rmd} , \quad
  \Bnq{\rmd} = \Bpq{\rmu} , \quad \text{and} \quad
  \Bnq{\rms} = \Bpq{\rms} .
\end{equation}
The $\pi$-nucleon sigma term, $\SigmapiN$, may be written as
\begin{equation} \label{eqn:SigmapiN}
  \SigmapiN \equiv \frac{1}{2} (\mup + \md)
                   \times \left( \BNq{\rmu} + \BNq{\rmd} \right) \; ,
\end{equation}
and the coefficients $\fTq{q}$ may be written in the forms:
\begin{alignat}{3}
  \label{eqn:fNTu}
  \fTq{\rmu}
    & \ = \ & \frac{\mup \Bq{\rmu}}{m_N}
    & \ = \ & \frac{2 \SigmapiN}{m_N (1+\frac{\md}{\mup})
                                 (1+\frac{\Bq{\rmd}}{\Bq{\rmu}})}
    \; , \\
  \label{eqn:fNTd}
  \fTq{\rmd}
    & \ = \ & \frac{\md \Bq{\rmd}}{m_N}
    & \ = \ & \frac{2 \SigmapiN}{m_N (1+\frac{\mup}{\md})
                                 (1+\frac{\Bq{\rmu}}{\Bq{\rmd}})}
    \; , \\
  \label{eqn:fNTs}
  \fTq{\rms}
    & \ = \ & \frac{\ms \Bq{\rms}}{m_N}
    & \ = \ & \frac{(\frac{\ms}{\md}) \SigmapiN \, y}%
                   {m_N (1+\frac{\mup}{\md})}
    \; ; \quad\quad\quad
\end{alignat}
where we have dropped the $(N)$ superscript from $\fTq{q}$ and
$\Bq{q}$.

The effects of the uncertainties in the $\fTq{q}$ were considered
in~\cite{priorvals}, and we were motivated to
study~\cite{Ellis:2005mb} variations in the value of $y$ by
re-evaluations of  the $\pi$-nucleon sigma term $\SigmapiN$, which is
related to the strange scalar density in the nucleon by
\begin{equation} \label{eqn:y2}
  y = 1 - \sigma_0/\SigmapiN \; .
\end{equation}
The value for $\sigma_0$ given in \reftab{params} is estimated on the
basis of octet baryon mass differences to be $\sigma_0 = 36 \pm 7$~MeV
\cite{Borasoy:1996bx,Gasser:1990ce,Knecht:1999dp,Sainio:2001bq}.
Based on recent determinations of $\SigmapiN$
at the Cheng-Dashen point~\cite{Pavan:2001wz},
one finds
\begin{equation} \label{eqn:SigmapiNexp}
  \SigmapiN \, = \, (64 \pm 8) \; \text{MeV} \; .
\end{equation}
We use the central value in \refeqn{SigmapiNexp} as our default,
namely $\SigmapiN = 64$~MeV.

\begin{figure*}
  \insertdoublefig{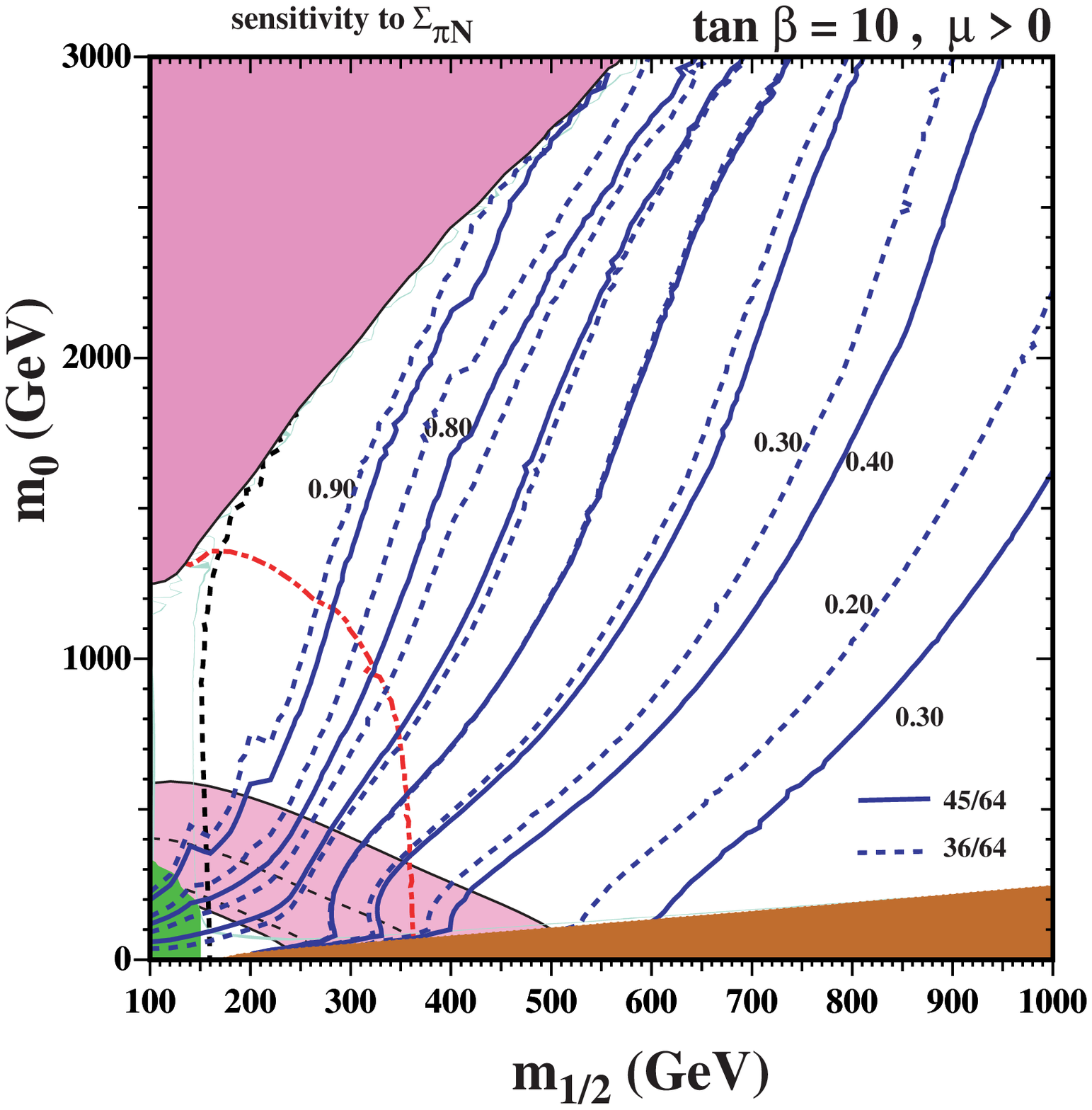}{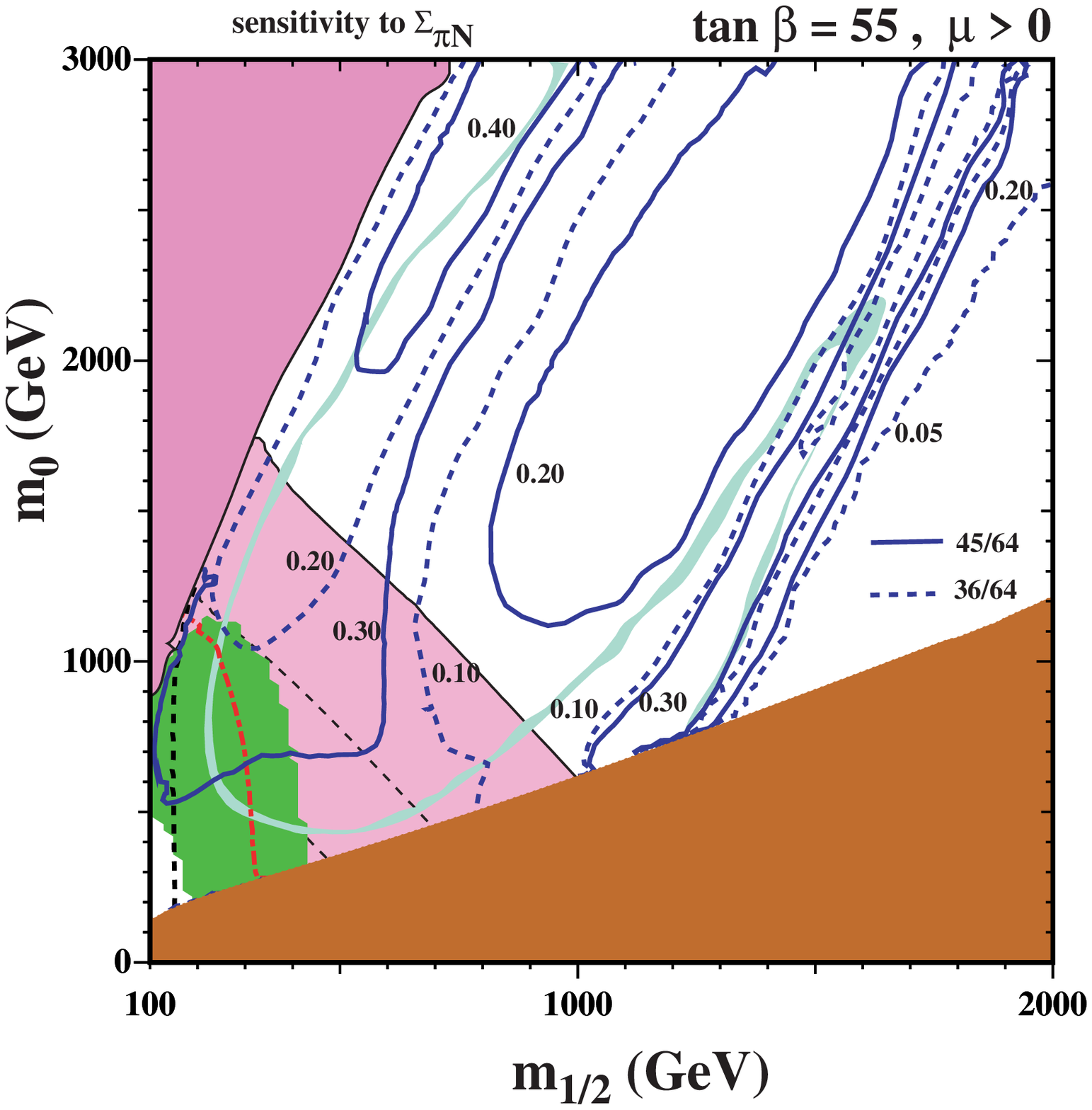}
  \caption{\it
    \coloronlinestatement
    The $(m_{1/2},m_0)$ planes for the CMSSM for $A_0 = 0$ and (left)
    $\tanb = 10$, (right) $\tanb = 55$, showing contours of the
    ratios of the solar dark matter annihilation rates calculated
    using $\Sigma_{\pi N} = 36$~MeV (dashed lines) and 45~MeV (solid
    lines) to calculations with the default value
    $\Sigma_{\pi N} = 64$~MeV.
    Also shown are the theoretical, phenomenological, experimental and
    cosmological constraints described in the text.
    }
  \label{fig:sigma}
\end{figure*}

In \reffig{sigma}, we compare the
annihilation rates calculated with this default value and with
$\SigmapiN = 45$ or 36~MeV, the latter corresponding via
\refeqn{y2} to the central value for $\sigma_0$, \ie, $y = 0$.
As could be expected from the previous analysis demonstrating
the importance of spin-independent scattering at large $m_{1/2}$,
we see that the annihilation rate decreases very substantially if
$\SigmapiN$ is decreased to 45 or 36~MeV~\footnote{The 
contours near the rapid-annihilation funnel
again reflect the non-equilibrium effects on the annihilation 
rate discussed earlier.}.
\textit{As in the case of direct dark matter detection via
spin-independent scattering, it is also very important for the
interpretation of indirect dark matter searches via neutrinos from
dark matter annihilation inside the Sun to pin down the magnitude of
$\SigmapiN$.}

\section{\label{sec:fluxes} Neutrino and Muon Fluxes}

Neutrino detectors such as IceCube/DeepCore \cite{Ahrens:2003ix,
Wiebusch:2009jf} aim to detect
the capture of LSPs in the Sun using neutrinos produced in the LSP
annihilations.  The primary experimental signature is the passage of
neutrino-induced muons through the detector.  Here, we examine the flux
of neutrinos and neutrino-induced muons in the $\tanb = 10$ and 55
planes, as well as along the WMAP-preferred focus-point,
coannihilation, and funnel-region strips (the last for $\tanb = 55$
only). We also study the neutrino and flux spectra for
several benchmark scenarios given in \reftab{models}.

Determining the neutrino spectrum in a detector involves two
steps: (1) the production of neutrinos from LSP annihilations, and
(2) propagation of the neutrinos from the interior of the Sun to the
detector.  Low energy neutralinos do not generally annihilate directly
into neutrinos; instead, neutrinos are produced in decays/showers
of the primary annihilation particles such as $W$ bosons, top quarks, or
tau leptons (\eg\ $\chi\chi \to \tau\bar{\tau}$, with
$\tau \to \mu \bar{\nu}_{\mu} \nu_{\tau}$).  Determination of the
neutrino spectra in such showers is complicated by the fact that they
occur in the dense center of the Sun, so that the primary particles may
lose energy before decaying~\cite{Ritz:1987mh,Giudice:1988vs,
Edsjo:1993pb,Jungman:1994jr}.

Once neutrinos are produced, they must travel
through the Sun, where they may undergo charged-current or
neutral-current interactions that absorb the neutrinos or reduce the
neutrino energies, respectively.  The Sun is mainly transparent to
neutrinos well below $\sim$100~GeV, but becomes opaque when the
energies reach $\sim$200-300~GeV; thus, high-energy neutrinos are
heavily suppressed.  Oscillations between neutrino species between
the Sun and Earth must also be taken into account~\cite{Ellis:1992df}.

We use the neutrino and neutrino-induced muon spectra determined by
WimpSim \cite{wimpsim} (and used within DarkSUSY~\cite{darksusy}),
which simulated both of the steps described above to generate spectra
for several LSP annihilation channels.  Neutrino production and
propagation has also been simulated in Ref.~\cite{Cirelli:flux}%
~\footnote{
  During the preparation of this paper, an error was discovered in
  the anti-neutrino fluxes of Ref.~\cite{Cirelli:flux} that prevented
  us from making full use of their results at this time.
  }.

For our neutrino spectrum results here, we neglect Higgs annihilation
channels and include only annihilations into quarks, leptons,
$W$ and $Z$ bosons.  The Higgs channels account for less than 3\% of the
annihilations in all the $\tanb = 10$ region we have considered
and, apart from a small region at $m_{1/2} > 1900$~GeV just above the
charged dark matter constraint (brown region in the planar
figures), Higgs channels account for less than 10\% of the
annihilations in the $\tanb = 55$ plane.  In most cases, and all of
the WMAP strips, the Higgs channels account for $\sim$1\% or less of
the annihilations.  The error introduced by neglecting the Higgs
channels is thus far smaller than variations induced by the choice
of solar model or uncertainties in the hadronic parameters discussed
previously.

The muon energy detection threshold $\Eth$ is an important
consideration in determining the sensitivity of neutrino detectors
to annihilations in the Sun.  Increasing the exposure
area of a detector often requires a sacrifice in the energy threshold.
Since high-energy neutrinos from the Sun (and thus muons) are
suppressed, high thresholds may inhibit detectors from observing
this type of signal.  The full IceCube detector will have a fairly
large effective area, but gradually loses sensitivity to muons below
energies $\sim$100~GeV due to the large spacing between
strings~\cite{Ahrens:2003ix}.  A smaller portion of the
detector, referred to as DeepCore, will be instrumented more densely
and with higher efficiency phototubes to allow for detection
of lower-energy muons; DeepCore can potentially detect muons down to
energies of 10~GeV or lower~\cite{Wiebusch:2009jf}.
The effective neutrino detection area of the combined IceCube/DeepCore
detector falls from $\sim$1~m$^2$ at 1~TeV, to $\sim$10$^{-2}$~m$^2$
at 100~GeV, to $\sim$10$^{-4}$~m$^2$ at 10~GeV, with IceCube providing
nearly all of the effective area above 100~GeV and DeepCore providing
nearly all of it below 30~GeV.

The ability of IceCube/DeepCore to observe a signal above background
for a particular CMSSM model depends on the neutrino spectrum, angular
spread of the induced muons, backgrounds, and detector
geometry~\cite{Resconi:2008fe,Abbasi:2009uz}.
The predominant backgrounds are muons and muon neutrinos generated
from cosmic ray interactions in the atmosphere; cosmic ray interactions
in the Sun may also contribute to the background \cite{Ingelman:1996mj}.
When looking for a signal from a point source such as the Sun, the
diffuse background is reduced by making angular cuts on the observed
muon event directions.  A signal analysis is complicated by the fact
that the angular spread of solar neutrino-induced muons is dependent
on neutrino energies, so that optimal data cuts are dependent on the
expected neutrino spectrum for a particular CMSSM model.  In addition,
the angular and energy resolutions and the effective area for muon
events, particularly at low energies, is dependent on the detector
geometry.  Such a careful analysis is beyond the scope of this paper.

Still, we can roughly approximate the IceCube/DeepCore sensitivity
to estimate the detectability of various CMSSM models.  IceCube/DeepCore
has generated a conservative sensitivity limit for muons above 1~GeV
in their detector
assuming a particular hard annihilation spectrum~\cite{Abbasi:2009uz}
($\tau \bar{\tau}$ for $m_{\chi} < 80$~GeV, $W^+ W^-$ at higher masses).
The fraction of annihilations into channels that yield hard spectra
($\chi\chi \to ZZ, W^+ W^-, t\bar{t}$ or $\tau\bar{\tau}$) is never
trivial in the CMSSM models we consider in this paper.
In the $\tanb = 10$ case, 70--99\% of the annihilations are into one of
these hard channels, whilst in the $\tanb = 55$ case 20--50\% of the
annihilations are into one of these channels. 
The fractions of annihilations into soft spectra channels
(\eg\ $\chi\chi \to b\bar{b}$) are 1--30\% and 50--80\% for the
$\tanb = 10$ and 55 cases, respectively.  Thus, in all cases, the
spectrum produced in a CMSSM model will contain 
a significant hard component.
Since these CMSSM neutrino spectra differ from the spectra used by
IceCube/DeepCore to generate their muon flux limits, the
IceCube/DeepCore muon flux limits cannot be directly applied to
particular CMSSM models.  However, the significant hard component of
the CMSSM spectrum should be similar enough to the hard spectrum
channels analyzed by IceCube/DeepCore that the sensitivity limits are
within a factor of 2 ($\tanb=10$) or 2--5 ($\tanb=55$) of the
sensitivity that would be obtained had the actual neutrino spectra been
used in the analysis.  Hence,
in the $\tanb=55$ case, the limits obtainable are expected
to be somewhat weaker than in the $\tanb=10$ case,
due to the lower fraction of annihilations into hard channels.

The IceCube/DeepCore projected sensitivity is shown in
Figures~\ref{fig:ThreshStrip}-\ref{fig:SigmaStrip}, normalized to the
total expected muon flux above 1~GeV (though the experiment is only
weakly sensitive to muons at such low energies) \footnote{
  If the actual spectrum were to be entirely soft, the IceCube/DeepCore
  muon flux sensitivity would be weakened by an order of magnitude or
  more.
  }.
We note that the IceCube/DeepCore collaboration characterizes this as
a ``conservative'' limit, and that the sensitivity could be further
improved in several possible ways. For example, (1) data are generally
used only from periods when the Sun is below the horizon, as this
reduces the cosmic-ray backgrounds.  However, it may be possible to use
the surrounding IceCube portion of the detector
to veto cosmic-ray-induced events in the DeepCore portion of the
detector, allowing data from daylight hours to be included in the
analysis and hence potentially doubling the exposure at low energies.
In addition, (2) even a few more strings placed in the DeepCore
region (a very interesting possibility in light of our analysis) would
significantly improve sensitivity to low-energy muons.
For these reasons, the IceCube/DeepCore muon flux sensitivity limits
should be taken only as an order of magnitude estimate when assessing
below the detectability of the CMSSM models below.

\subsection{Fluxes in $(m_{1/2}, m_0)$ Planes}

\begin{figure*}
  \insertdoublefig[0.40]{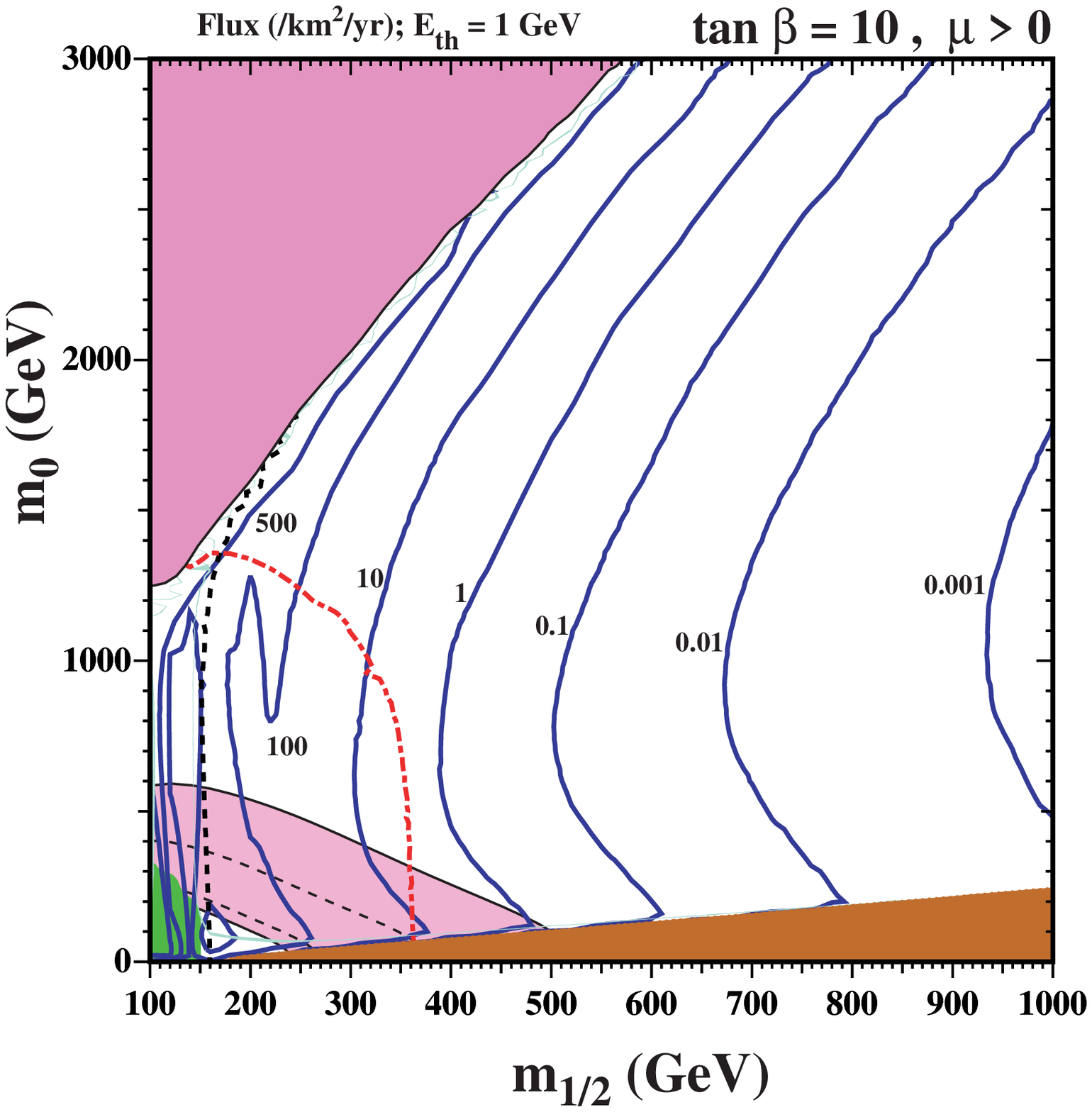}%
                        {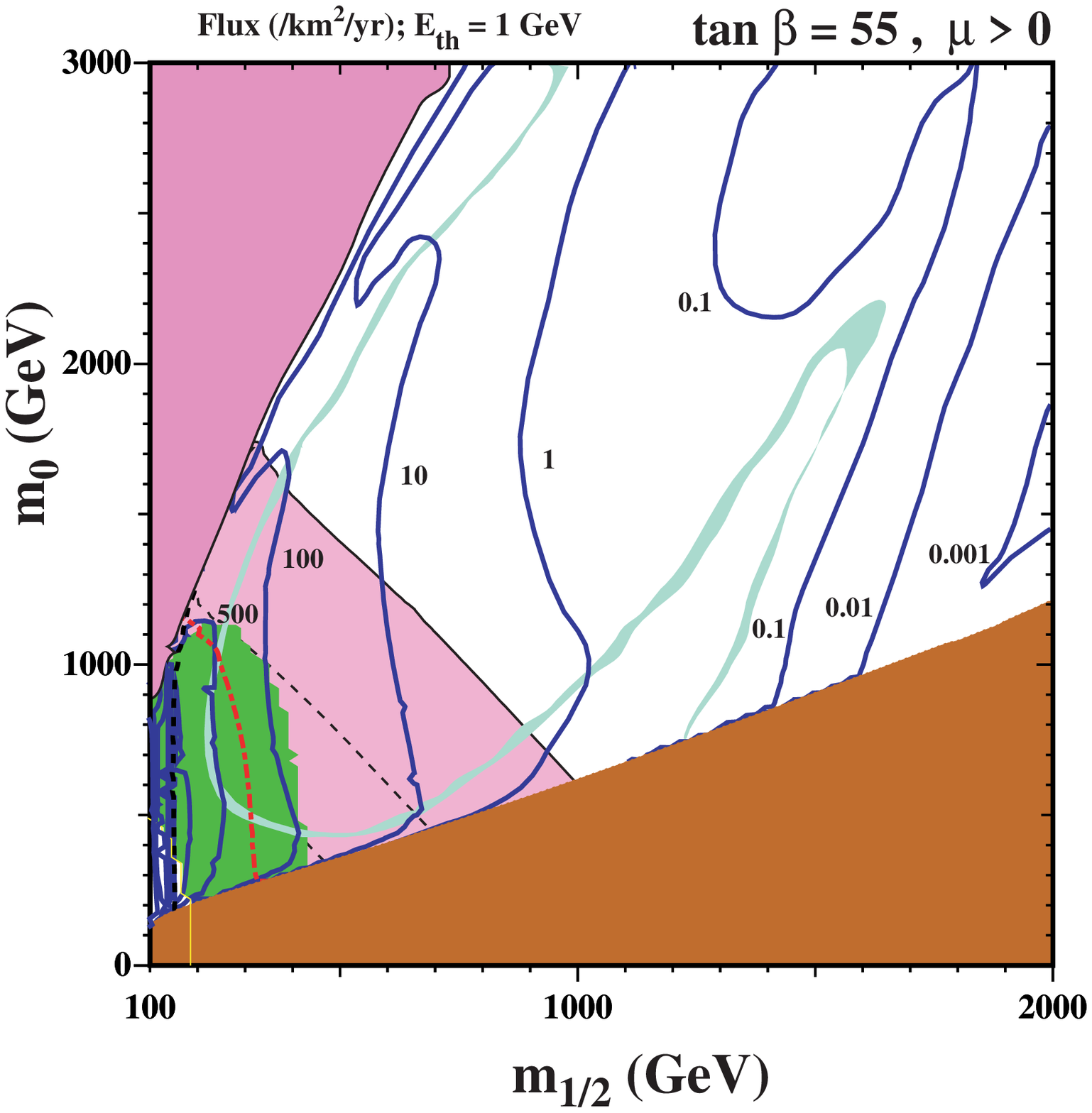}\\
  \insertdoublefig[0.40]{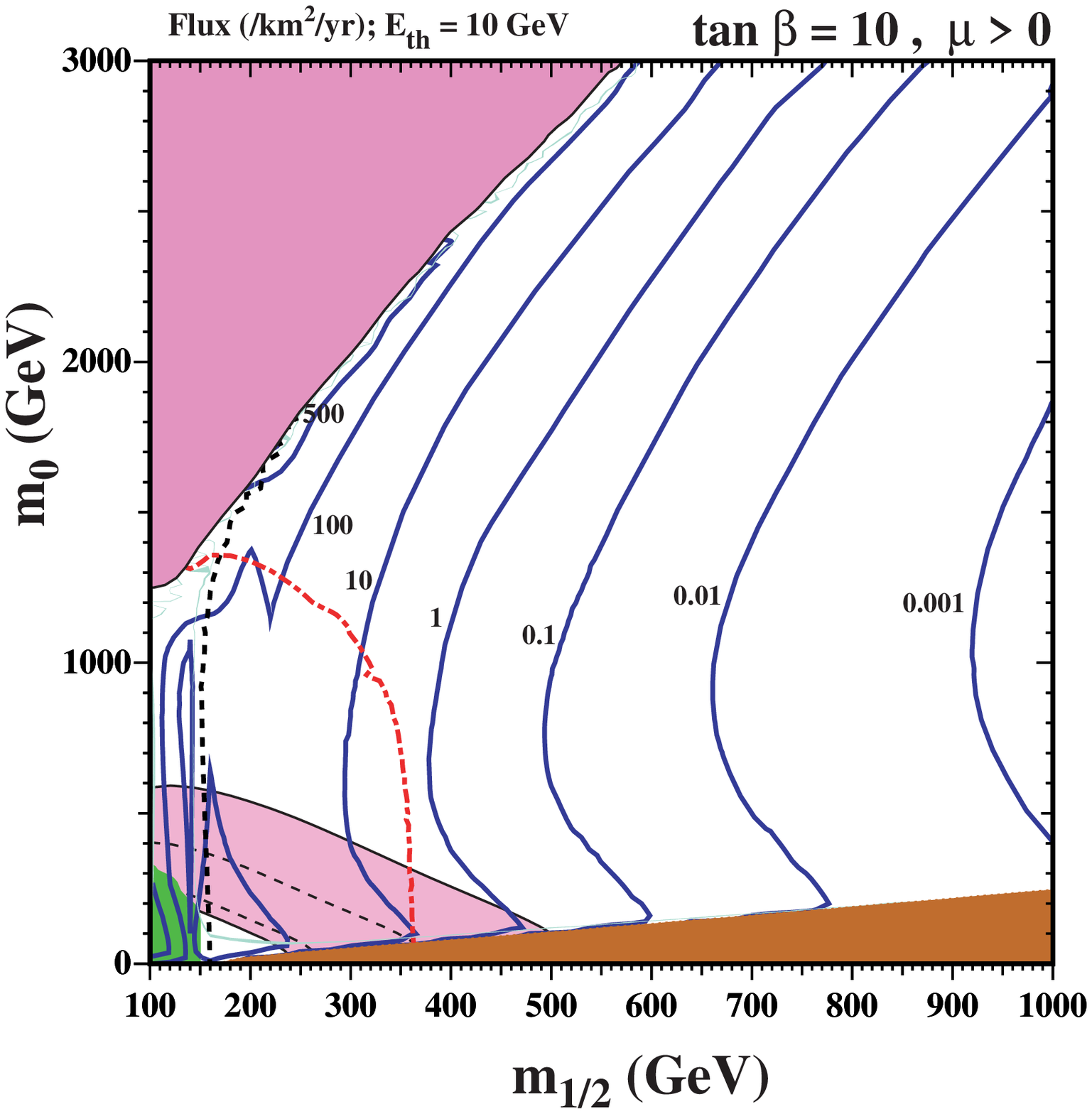}%
                        {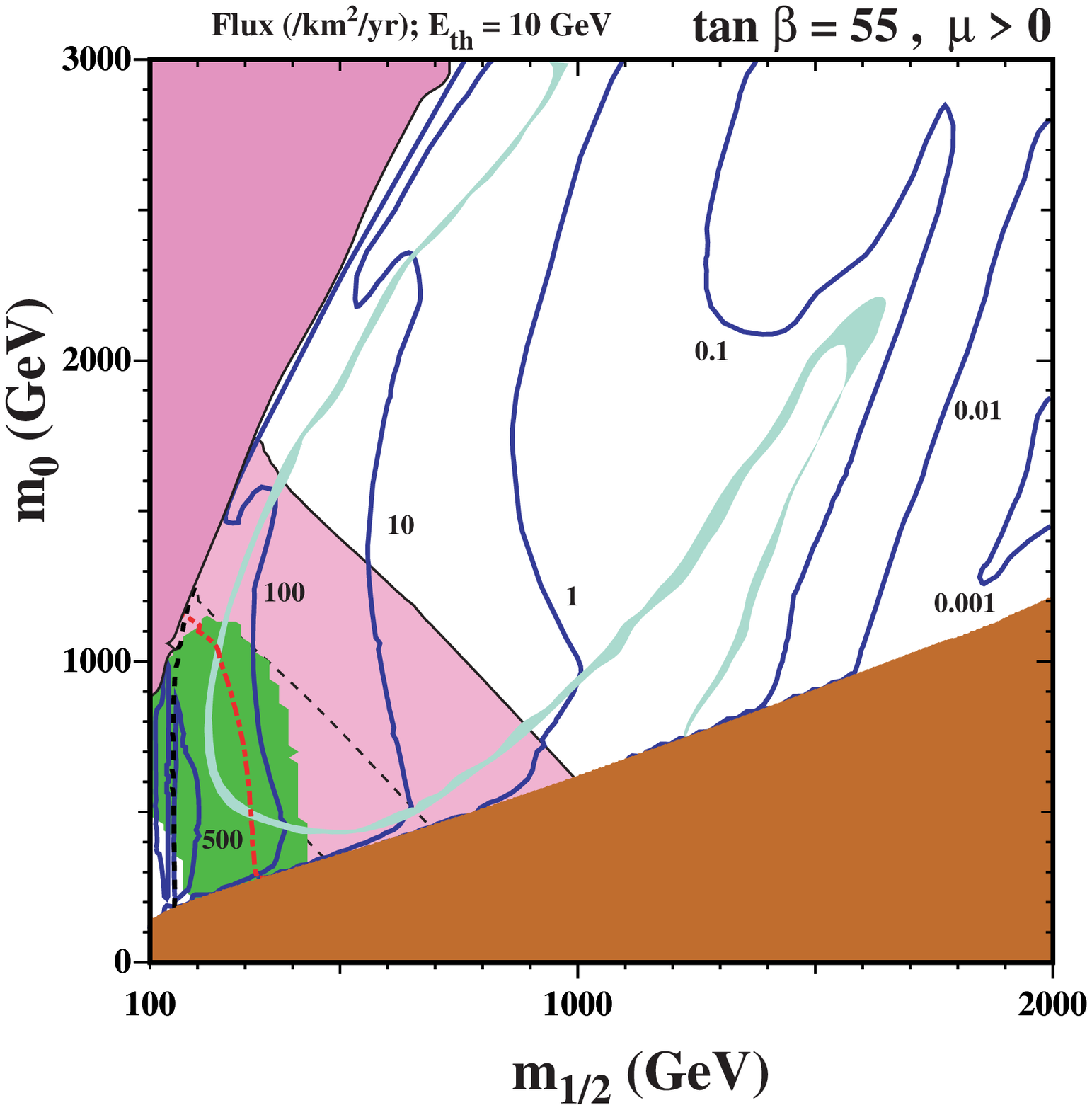}\\
  \insertdoublefig[0.40]{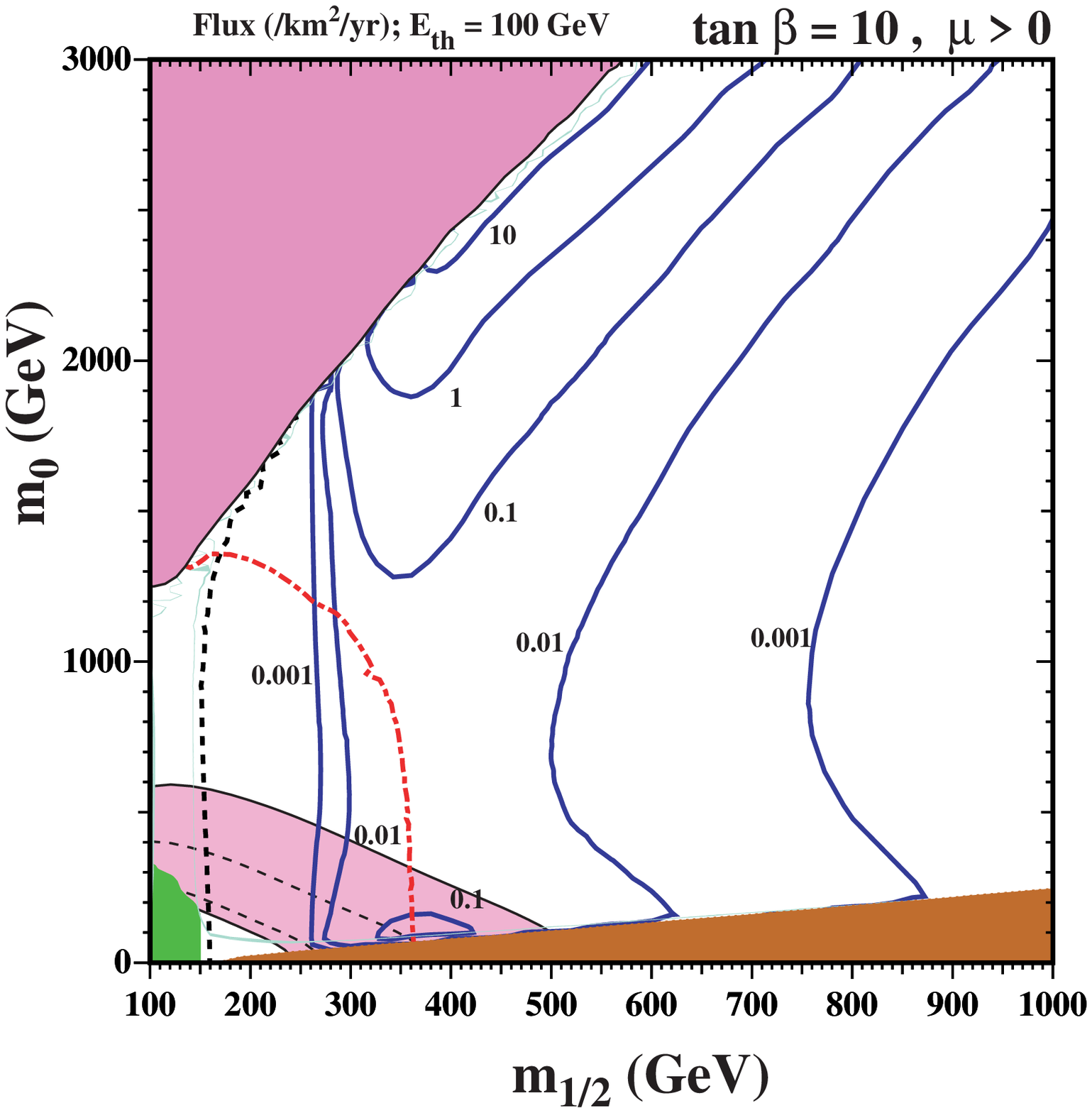}%
                        {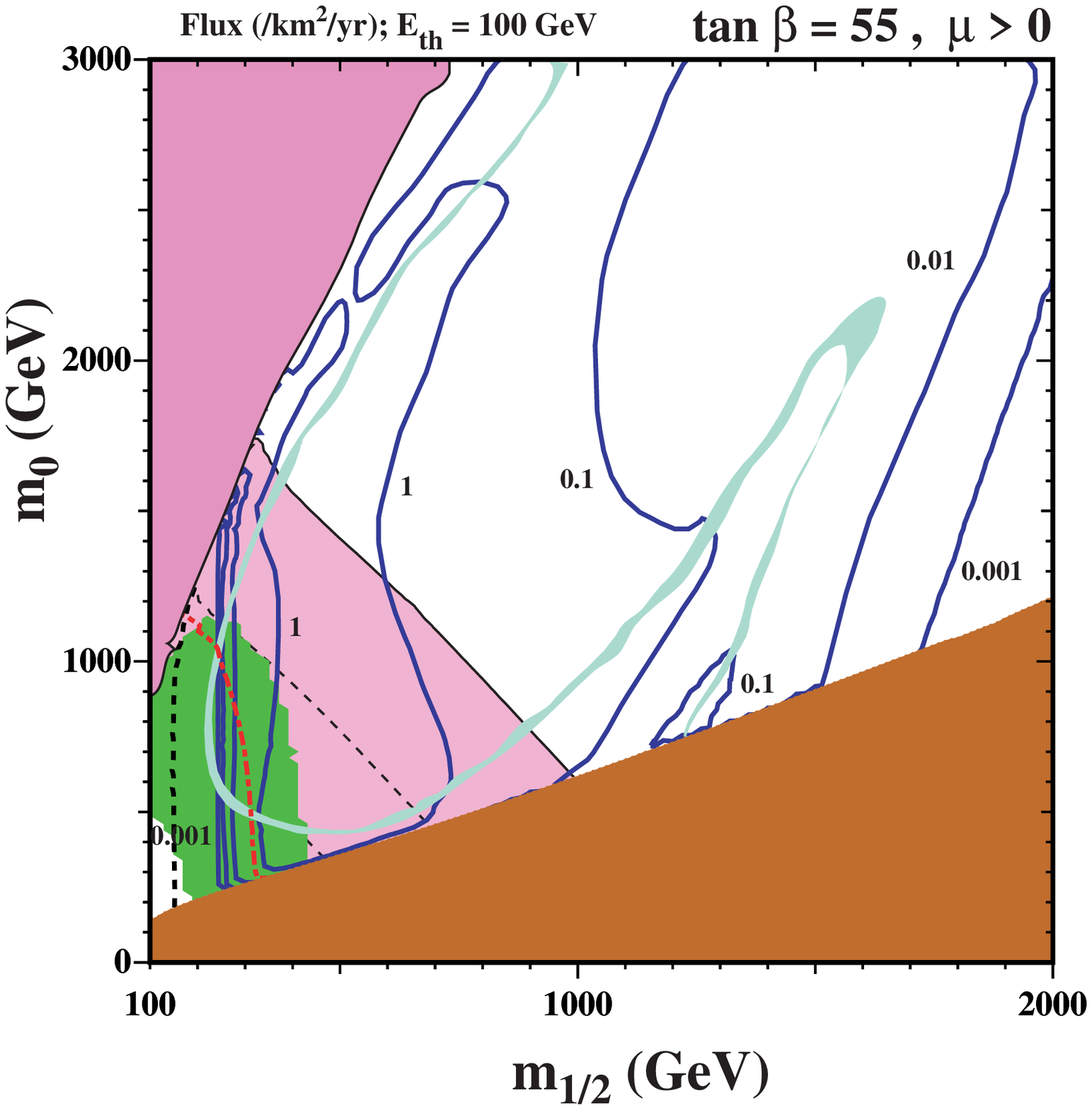}
  \caption{\it
    \coloronlinestatement
    The $(m_{1/2},m_0)$ planes for the CMSSM for $A_0 = 0$ and (left)
    $\tanb = 10$, (right) $\tanb = 55$, showing contours of the
    neutrino-induced muon flux through a detector with
    thresholds of 1~GeV (top), 10~GeV (middle), and 100~GeV (bottom).
    Also shown are the theoretical, phenomenological, experimental and
    cosmological constraints described in the text.
    }
  \label{fig:Thresh}
\end{figure*}

We present in \reffig{Thresh} contours of the
neutrino-induced muon fluxes in the  $(m_{1/2}, m_0)$ planes for
$\tanb = 10$ (left) and $\tanb = 55$ (right), for three 
values of the detector threshold: 1~GeV (top), 10~GeV (middle), and
100~GeV (bottom).
The other phenomenological, experimental and  cosmological constraints
are as in \reffig{capann}.
As the aforementioned sensitivity limits cannot easily be applied to
these figures, we can expect, to a very rough approximation, IceCube to
detect muon fluxes on the order of 10 or 10$^2$~/km$^2$/yr above
$\sim$100~GeV and DeepCore to detect muon fluxes
on the order of 10$^2$ or 10$^3$~/km$^2$/yr above $\sim$10~GeV.

For $\tanb = 10$, the muon flux attains its maximum value close to the
focus-point region.  The picture does not change much when the detector
threshold increases from 1 to 10~GeV, but for threshold 100~GeV the
contours show lower muon fluxes.
IceCube/DeepCore is in this case expected to probe all the
focus-point region and much of the bulk region. Note that although the
highest flux contour we display is 500 /km$^2$/yr, the flux continues
to increase as one approaches the region where radiative electroweak
symmetry breaking no longer is viable. The focus-point strip is very
close to this boundary and indeed the flux does exceed
$10^3$ /km$^2$/yr when the relic density is in the WMAP range.
This will be seen more clearly when we discuss the the WMAP strips in
the next subsection.
The focus-point region produces the largest fluxes observable by
IceCube alone (high threshold, bottom left panel of \reffig{Thresh});
these high-energy muon fluxes may even be detectable without including
the sensitivity to lower energy muons that DeepCore provides.
The same holds for
$\tanb = 55$: the neutrino-induced muon flux is again largest in the
focus-point region, close to the region theoretically excluded due to
the absence of electroweak symmetry breaking.
However, the fluxes are likely too low to be observed by
IceCube/DeepCore, except for the lowest values of $m_{1/2}$ along
this strip.
The cosmologically-favoured region around the rapid-annihilation funnel
yields much smaller values for the muon fluxes, below the expected
experimental sensitivities.

\subsection{Fluxes along WMAP Strips}

\begin{figure*}
  \insertwidefig{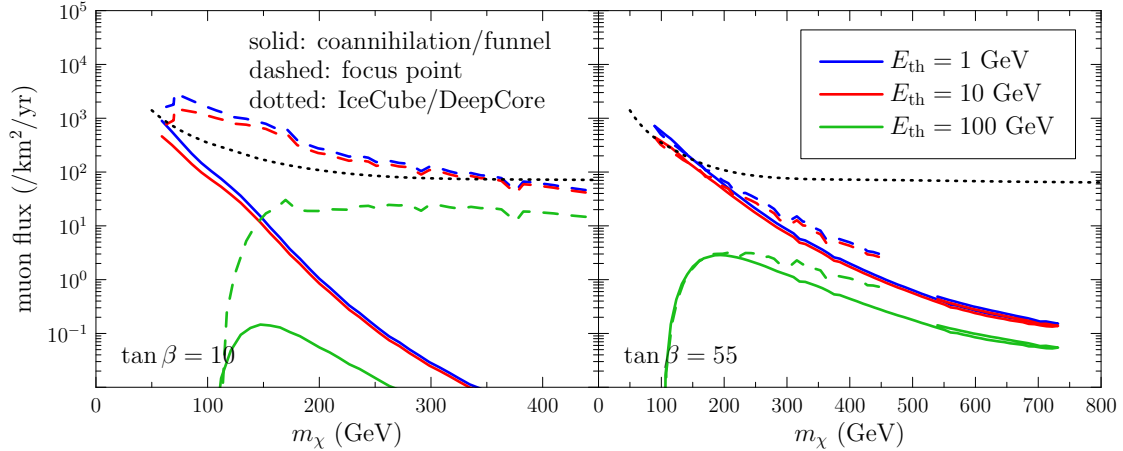}
  \caption{\it
    \coloronlinestatement
    The CMSSM muon fluxes though a detector calculated for $A_0 = 0$
    and (left) $\tanb = 10$, (right) $\tanb = 55$,
    along the WMAP strips in the coannihilation/funnel regions (solid)
    and the focus-point region (dashed).
    Fluxes are shown for muon energy thresholds of (top to bottom)
    1~GeV, 10~GeV, and 100~GeV.
    Also shown is a conservative estimate of sensitivity of the
    IceCube/DeepCore detector (dotted), normalized to a muon threshold
    of 1~GeV, for a particular hard annihilation spectrum
    ($\tau \bar{\tau}$ for $m_{\chi} < 80$~GeV, $W^+ W^-$ at higher
    masses).
    The IceCube/DeepCore sensitivity shown does not directly apply
    to the CMSSM flux curves, but can be treated as a rough
    approximation to which CMSSM models might be detectable,
    as discussed in the text.
    }
  \label{fig:ThreshStrip}
\end{figure*}

In \reffig{ThreshStrip}, we plot the neutrino-induced muon fluxes in the  
cosmologically-favored regions of the CMSSM that are also compatible
with the rest of the phenomenological and experimental constraints,
namely the regions that we call WMAP strips. 
These strips are determined by fixing the value of $m_0$ so that at
each value of $m_{1/2}$ (or equivalently $m_\chi \approx 0.43 m_{1/2}$)
the relic density is within the 2 $\sigma$ WMAP range
\cite{Komatsu:2008hk}.  We pick again the same
representative values of $\tanb =10$ (left) and $55$ (right).
For $\tanb = 10$, there are two such strips corresponding to either 
the co-annihilation or focus-point regions.  For $\tanb = 55$,  the
co-annihilation strip runs into the funnel region at large $m_{\chi}$.
At very large $m_{\chi}$, we have sampled both sides of the funnel
resulting in a `doubling back' of the curves we display.

A conservative IceCube/DeepCore sensitivity limit for muon fluxes above
1~GeV is also shown in the figure.  As discussed previously, this
sensitivity cannot be directly applied to the CMSSM fluxes shown as
the spectra differ.  However, the limit should be within a factor of
2 or so of the muon fluxes necessary to detect these CMSSM models
(recall that, if additional strings are added to DeepCore and/or
data can be used from periods when the Sun is above the horizon, the
IceCube/DeepCore sensitivity could be as much as an order of magnitude
better than shown).

As observed in \reffig{Thresh}, the muon flux is largest in the
focus-point region for both values of $\tanb$. In \reffig{ThreshStrip}
(left) we plot the values for the muon fluxes along the coannihilation
strip (solid curve) and the focus-point strip (dashed curve) for
$\tanb=10$. The blue, red and green curves correspond to threshold
energies 1, 10 and 100~GeV. In the focus-point region, the muon flux
is $\sim 10^2$ -- $10^3$~/km$^2$/yr, making detection in
IceCube/DeepCore a likely possibility, particularly for lighter LSPs
(smaller $m_{1/2}$).
For $\tanb=55$ (right plot), the solid curve represents both the
coannihilation and funnel strips.  The double covering of
$m_{\chi} \sim 550-750$~GeV is due to the shape of the WMAP strip in
the funnel region, as noted above.
As already observed, the values of the muon fluxes are similar in the
coannihilation/funnel region and in the focus-point region for
$\tanb = 55$, but are smaller than those for $\tanb=10$, especially in
the focus-point region.

\begin{figure*}
  \insertwidefig{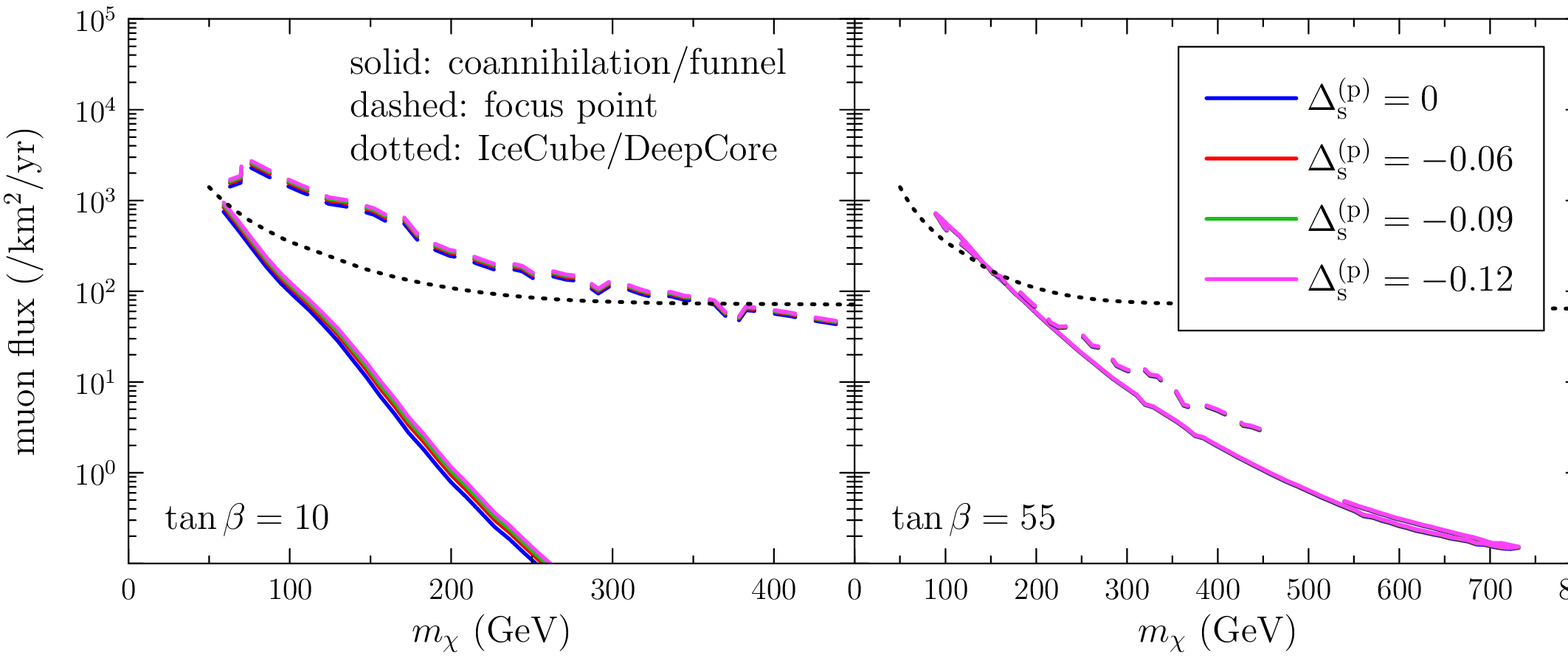}
  \caption{\it
    \coloronlinestatement
    The CMSSM muon fluxes though a detector calculated for $A_0 = 0$
    and (left) $\tanb = 10$, (right) $\tanb = 55$,
    along the WMAP strips in the coannihilation/funnel regions (solid)
    and the focus-point region (dashed). The fluxes are shown for
    $\Deltaps$ values of $0$, $-0.06$, $-0.09$, and $-0.12$
    with a detector threshold of 1~GeV.
    Also shown is a conservative approximation of sensitivity of the
    IceCube/DeepCore detector (dotted), as described in
    \reffig{ThreshStrip} and the text.
    }
  \label{fig:DeltasStrip}
\end{figure*}

\begin{figure*}
  \insertwidefig{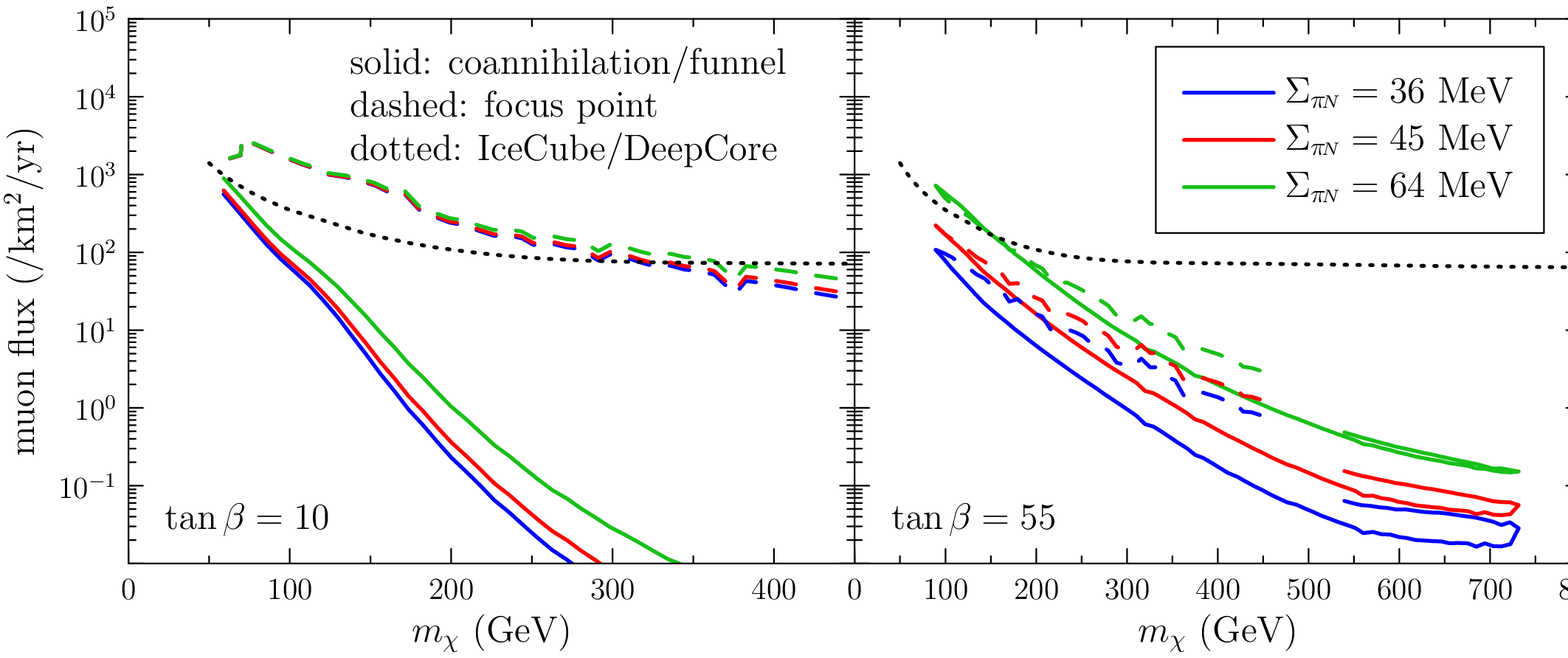}
  \caption{\it
    \coloronlinestatement
    The CMSSM muon fluxes though a detector calculated for $A_0 = 0$
    and (left) $\tanb = 10$, (right) $\tanb = 55$,
    along the WMAP strips in the coannihilation/funnel regions (solid)
    and the focus-point region (dashed). The fluxes are shown for 
    $\SigmapiN$ values of (top to bottom)
    $64$, $45$, and $36$~MeV
    with a detector threshold of 1~GeV.
    Also shown is a conservative approximation of sensitivity of the
    IceCube/DeepCore detector (dotted), as described in
    \reffig{ThreshStrip} and the text.
    }
  \label{fig:SigmaStrip}
\end{figure*}

We see in \reffig{DeltasStrip} that the muon fluxes are quite
insensitive to the value of $\Deltaps$ within the (plausible) range
studied. On the other hand, \reffig{SigmaStrip} shows that the muon
fluxes are more sensitive to the value of $\SigmapiN$, particularly for
$\tanb = 55$ and along the coannihilation strip for $\tanb = 10$ at
larger $m_{1/2}$. The comparisons shown are for a threshold energy of
1~GeV, but similar conclusions apply for thresholds of 10 and 100~GeV
(not shown).  The value of $\SigmapiN$ clearly impacts the ability of
IceCube/DeepCore to detect $\tanb = 55$ models.
 
\subsection{Fluxes for Benchmark Points}

\begin{table*}
  \begin{ruledtabular}
  \begin{tabular}{lcccc}
    Model             & C    & E    & L    & M    \\
    \hline
    $m_{1/2}$ (GeV)   & 400  & 300  & 460  & 1075 \\
    $m_{0}$ (GeV)     & 96   & 2003 & 312  & 1045 \\
    $\tanb$           & 10   & 10   & 50   & 55   \\
    $A_0$             & 0    & 0    & 0    & 0    \\
    $\signmu$         & +    & +    & +    & +    \\
    \hline
    $m_{\chi}$ (GeV)  & 165  & 117  & 193  & 474  \\
    \hline
    capture rate (/s)
            & $2.58 \times 10^{20}$
            & $4.97 \times 10^{22}$
            & $1.05 \times 10^{21}$
            & $9.79 \times 10^{18}$ \\
    annihilation rate [2$\Gamma$] (/s)
            & $8.72 \times 10^{19}$
            & $4.97 \times 10^{22}$
            & $1.05 \times 10^{21}$
            & $8.75 \times 10^{18}$ \\
    \hline
    Neutrino flux (/km$^2$/yr)  & & & & \\
    $\quad$$\Eth = 1$~GeV
            & $3.27 \times 10^{9}$
            & $9.19 \times 10^{11}$
            & $3.33 \times 10^{10}$
            & $2.16 \times 10^{8}$ \\
    $\quad$$\Eth = 10$~GeV
            & $2.60 \times 10^{9}$
            & $7.03 \times 10^{11}$
            & $2.43 \times 10^{10}$
            & $1.47 \times 10^{8}$ \\
    $\quad$$\Eth = 100$~GeV
            & $2.86 \times 10^{8}$
            & $8.80 \times 10^{9}$
            & $3.26 \times 10^{9}$
            & $3.35 \times 10^{7}$ \\
    \hline
    Muon flux (/km$^2$/yr)  & & & & \\
    $\quad$$\Eth = 1$~GeV
            & 5.98
            & 1210
            & 61.8
            & 0.830 \\
    $\quad$$\Eth = 10$~GeV
            & 4.71
            & 898
            & 49.7
            & 0.734 \\
    $\quad$$\Eth = 100$~GeV
            & 0.129
            & 0.243
            & 2.70
            & 0.222 \\
  \end{tabular}
  \end{ruledtabular}
  \caption[Benchmark models]{\it
    CMSSM parameters and results for benchmark models C, E, L, and M
    of~\protect\cite{bench}.
    $\Eth$ is the neutrino/muon energy threshold.
    Rates and fluxes are determined using the AGSS09 solar model and
    the central values found in \reftab{params} for the hadronic
    parameters.
    }
  \label{tab:models}
\end{table*}

We now display complete spectra for the CMSSM benchmark scenarios
C, E, L and M of~\cite{bench}. These are representative, respectively,
of the low-mass region of the coannihilation strip, the focus-point
region, the coannihilation region at larger $m_{1/2}$ and $\tanb$, and
the rapid-annihilation funnel region.
The benchmark parameters, capture/annihilation rates, and fluxes
are given in \reftab{models}.  Models E, L, and M are in or nearly in
equilibrium between capture and annihilation, whereas model C is
somewhat out of equilibrium.

\begin{figure}
  \insertfig{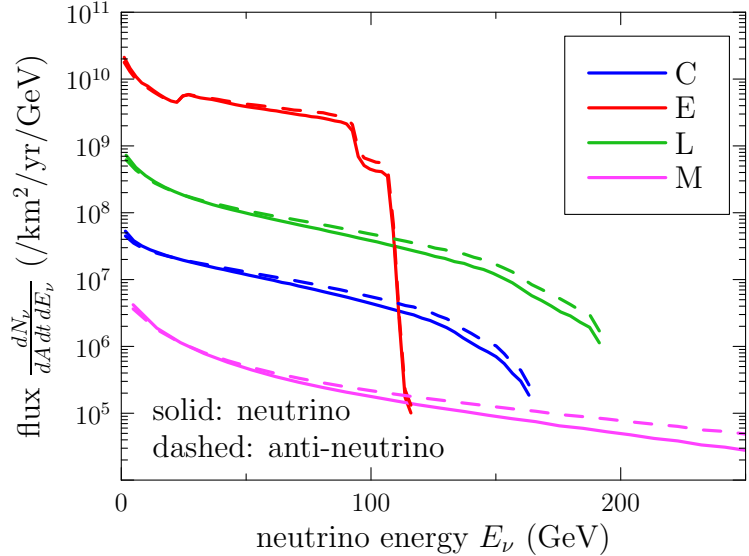}
  \caption{\it
    \coloronlinestatement
    The neutrino spectra in a terrestrial detector for (top to bottom)
    benchmark scenarios E, L, C, and M~\protect\cite{bench}.
    Neutrinos (solid) and anti-neutrinos (dashed) are shown separately.
    }
  \label{fig:nuSpectra}
\end{figure}

The differential neutrino fluxes for these scenarios are shown in
\reffig{nuSpectra}.  The highest differential flux is found in
model~E, partly because it has a larger annihilation rate than the
other models and partly because its neutrino spectrum is spread
over a smaller range of energies (at 117~GeV, this model has the
smallest LSP mass of the four).  Model~M has the lowest differential
flux, but the spectrum is spread out over a wider range of energies;
indeed, the spectrum for this model continues beyond the region shown.
In all four models, the spectra end at the corresponding LSP mass.
The plateau seen in model~E arises from $W \to \mu \nu_{\mu}$ decays
in the dominant $\chi\chi \to WW$ annihilation channel.  Since
the LSP mass is not much higher than the $W$ mass, the $W$s are produced
with relatively low speeds.  In this case, the neutrinos from the $W$
decays are kinematically limited to fall within the energy range
corresponding to the plateau in the spectrum.

\Reffig{nuSpectra} shows that the neutrino flux (solid) is slightly
smaller than the anti-neutrino flux (dashed), particularly at higher
energies.  This arises from the higher neutral- and charged-current
scattering cross sections for neutrinos, compared to anti-neutrinos,
when passing through the Sun.  The difference between neutrino and
anti-neutrino fluxes grows more pronounced at higher energies. 
However, both neutrinos and anti-neutrinos are
suppressed at higher energies, due to the increasing opacity of the
Sun: neutrinos are just more suppressed than anti-neutrinos.

\begin{figure}
  \insertfig{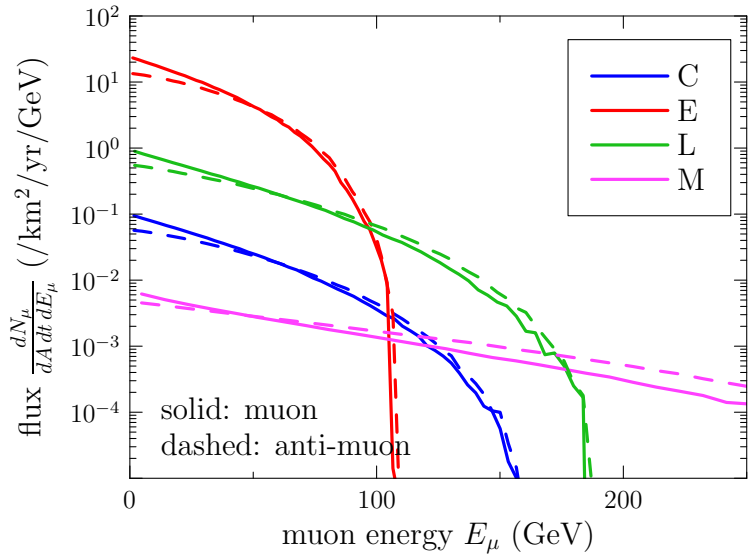}
  \caption{\it
    \coloronlinestatement
    The muon spectra in a terrestrial detector for (top to bottom)
    benchmark scenarios E, L, C, and M~\protect\cite{bench}.
    Muons (solid) and anti-muons (dashed) are shown separately.
    }
  \label{fig:muSpectra}
\end{figure}

\Reffig{muSpectra} shows the differential muon fluxes from these
neutrino spectra.  Model~E still has the highest differential flux,
but model~M has increased relative to the other models over the
intermediate energies shown in the figure, as compared to
\reffig{nuSpectra}.  This is because the higher-energy neutrinos
in this model are capable of inducing additional muons at a
range of lower energies.

Whilst model~E has the largest muon flux, it is all at relatively low
energies that IceCube alone cannot detect.  With a flux of just
under 10$^3$~/km$^2$/yr above 10~GeV (see \reftab{models}), however,
its flux lies above the expected DeepCore detection level.  Model~L
has the next highest spectrum, but would generate
$\sim$50~muons/km$^2$/yr above 10~GeV, probably falling
just below the DeepCore detection level.
However, its spectrum extends up to 190~GeV, yielding a significant
flux above 100~GeV of 2.7~/km$^2$/yr.  While not high enough to be
detected by IceCube alone, it is not far below the expected IceCube
detection level either.  The combination of DeepCore's low energy
sensitivity and IceCube's large volume make IceCube/DeepCore perhaps
able to detect this model.
Model~M, while producing high-energy muons to which
IceCube is most sensitive, has fluxes that are too low to be detected.
Model~C has neither the total flux nor an energy spectrum favorable for
detection.

\section{\label{sec:conclusions} Conclusions}

We have studied in detail the observability of muons produced by
neutrinos and antineutrinos generated by the annihilations of CMSSM
LSPs trapped inside the Sun, focusing in particular on models lying
along the parameter strips where the relic LSP density falls within the
range favoured by WMAP  \cite{Komatsu:2008hk}.  

For various reasons, discussed in \refsec{fluxes}, it is not
straightforward to apply direct detectability limits for
IceCube/DeepCore experiments on  the CMSSM parameter space.
The main complication is that the ability of the experiment to observe
a signal above background depends on the shape of the neutrino spectrum.
Nevertheless, a conservative estimation of the IceCube/DeepCore
detectability limits can be made, that being 10$^2$ to 10$^3$~/km$^2$/yr
for the muon flux above 1~GeV.
Based on this, our conclusions are not very encouraging:
along the coannihilation strips,
we find that only models near the low-$m_{1/2}$ ends of the strips for
$\tanb = 10$ and 55 may be detectable in IceCube/DeepCore.
Nearly all of the focus-point strip may be detectable by
IceCube/DeepCore for $\tanb = 10$; some of this strip may actually be
detectable with the bulk (\ie\ non-DeepCore) IceCube detector region
alone.
For $\tanb = 55$, IceCube/DeepCore will only be sensitive to the
low-$m_{1/2}$ end of the focus-point region.

As for the benchmark scenarios, we find that model~E
(from the focus-point region) is expected to be detectable in
IceCube/DeepCore, mainly due to the low energy muon sensitivity of
the DeepCore component as this model produces few high energy muons
that IceCube alone is sensitive to.
On the other hand, the muon spectrum in model L (in the coannhilation
region) extends to an energy of 190~GeV, and both the DeepCore and 
bulk IceCube components contribute to making this model borderline
detectable.
Model M (in the rapid-annihilation region), while also producing
high-energy muons, produces too few to be detectable in
IceCube/DeepCore, and model C (near the low-$m_{1/2}$ end of the
coannihilation strip) has neither the flux nor the energy spectrum
suitable for detection.
These conclusions are quite robust with respect to uncertainties in the
solar model. However, we do note that the calculated muon fluxes are
quite sensitive to the assumed value of the $\pi$-nucleon $\sigma$ term
$\SigmapiN$, which controls the magnitude of spin-independent LSP
scattering: we repeat our previous plea~\cite{Ellis:2008hf} to our
experimental colleagues to reduce the uncertainty in $\SigmapiN$.

These studies indicate that the indirect search for dark matter via
annihilations that yield high-energy neutrinos and hence muons may not
be the most promising way to discover supersymmetry, at least within the
restrictive CMSSM framework. On the other hand, there are representative
models where the calculated fluxes are close to the sensitivities of
DeepCore and/or IceCube. Therefore, if other evidence for superymmetry
were to be found, \eg, either at the LHC or in the direct search for
dark matter scattering, it would be interesting to increase the
sensitivities of DeepCore and IceCube, \eg, by decreasing the threshold
or by increasing the effective surface area.
\textit{Une affaire \`a suivre!}


\begin{acknowledgments}
  The work of KAO was supported in part by DOE Grant
  No.\ DE-FG02-94ER-40823.
  CS acknowledges the support of the William I.\ Fine Theoretical
  Physics Institute at the University of Minnesota and
  financial support from the Swedish Research Council (VR) through
  the Oskar Klein Centre.
  CS thanks A.~Heger and P.~Scott for discussions about solar models,
  J.~Edsj\"o for discussions about neutrino fluxes,
  and C.~Finley for discussions about the IceCube/DeepCore detector.
  The work of V.C.S. was supported by Marie Curie International
  Reintegration grant SUSYDM-PHEN, MIRG-CT-2007-203189.
\end{acknowledgments}




\end{document}